\documentclass[prl, twocolumn, superscriptaddress, nofootinbib, nobibnotes]{revtex4-2}

\usepackage[english]{babel}
\usepackage{hyperref}

\usepackage{soul}

\usepackage[pdftex]{graphicx}
\usepackage{color}
\usepackage[toc,page]{appendix}

\usepackage{amsmath}
\usepackage{amssymb}
\usepackage{mathtools}
\usepackage{braket}

\usepackage[sort&compress]{natbib}

\newcommand{\un}[1]{\ensuremath{\,\mathrm{#1}}}
\renewcommand{\v}[1]{\ensuremath{\boldsymbol{#1}}}

\newcommand{\fig}[1]{Figure~\ref{fig:#1}}
\newcommand{\Fig}[1]{Figure~\ref{fig:#1}}

\newcommand{\eq}[1]{(\ref{eq:#1})}
\newcommand{\lr}[1]{\ensuremath{\left( #1 \right)}}

\newcommand{\I}{\mathrm{i}}

\newcommand{\Sg}{\Sigma}
\newcommand{\sg}{\sigma}

\newcommand{\mc}{\mathcal}

\newcommand{\Tr}[1]{\ensuremath{\mathrm{Tr}\left(1\right)}}
\newcommand{\pd}{\partial}

\newcommand{\eps}{\varepsilon}
\newcommand{\bt}{\beta}
\newcommand{\dl}{\delta}
\newcommand{\abs}[1]{\left| #1 \right|}

\usepackage{ulem}

\begin{document}

\title{Graphene nanodrums as valleytronic devices}

\author{Walter Ortiz}
\email{balter20@gmail.com}
\affiliation{Instituto de Investigaci\'on en Ciencias B\'asicas y
  Aplicadas, Universidad Aut\'onoma del Estado de Morelos,
  Cuernavaca, Mexico}
\affiliation{Instituto de Ciencias F\'isicas, Universidad Nacional
  Aut\'onoma de M\'exico, Cuernavaca, Mexico}

\author{Nikodem Szpak}
\email{nikodem.szpak@uni-due.de}
\affiliation{Fakult\"at f\"ur Physik, Universit\"at Duibsurg-Essen,
  Duisburg, Germany}

\author{Thomas Stegmann}
\email{stegmann@icf.unam.mx}
\affiliation{Instituto de Ciencias F\'isicas, Universidad Nacional
  Aut\'onoma de M\'exico, Cuernavaca, Mexico}

\date{\today}

\begin{abstract}
We investigate the electronic transport in graphene nanoelectromechanical resonators (GrNEMS), known also as graphene nanodrums or nanomembranes. We demonstrate that these devices, despite small values of out-of-plane strain, between $0.1$ and $1\%$, can be used as efficient and robust valley polarizers and filters. Their working principle is based on the pseudomagnetic field generated by the strain of the graphene membrane.
They work for ballistic electron beams as well as for strongly dispersed ones and can be also used as electron beam collimators due to the focusing effect of the pseudomagnetic field. We show additionally that the current flow can be estimated by semiclassical trajectories which represent a computationally efficient tool for predicting the functionality of the devices.
\end{abstract}

\maketitle

\section{Introduction}

Graphene, the wonder material of the 21st century not only features pseudo-relativistic electrons with conical dispersion relation, but also facilitates two valleys located at the $\v{K}^+$ and $\v{K}^-$ points of the Brillouin zone. This degree of freedom, interpreted as the valley spin of the electrons, suggests a new kind of electronics, named valleytronics \cite{Schaibley2016}, where the valley spin is processed instead of the charge or the real spin. In recent years, several proposals have been made to create, manipulate and detect valley polarized currents in graphene \cite{Wang2014, Settnes2016, Stegmann2019, Milovanovic2016, Carrillo-Bastos2016, Zhai2018, Carrillo-Bastos2018, Mahmud2020, Solomon2021} and even the idea of valley-transistors has been put forward \cite{Lee2012}. However, valleytronics is still in its infancy as it remains challenging to control the valley polarized currents \cite{Faria2020}.

\begin{figure}[t]
  \centering
  \includegraphics[scale=0.39]{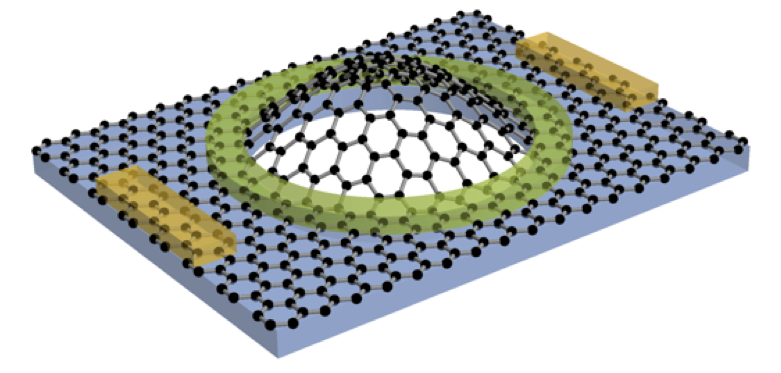}
  \caption{Sketch of the investigated device. A graphene membrane is deposited on an insulating substrate with a circular cavity (blue shaded region). The graphene membrane is clamped at the edges using, for example, photoresist (green shaded ring) forming a nanodrum. The graphene membrane is deformed by the pressure of an external gas or the electric field of metallic gates (not shown). Current is injected and detected at the edges of the system (golden bars).}
  \label{fig:1}
\end{figure}

In this paper, we show that graphene nanoelectromechanical systems (GrNEMS) \cite{Chen2013b, Castellanos-Gomez2014, Khan2017, Lemme2020} can be used to construct efficient and robust valley polarizers. We consider a graphene membrane that is spanned over a cavity, as shown in \fig{1}. The membrane can be deformed in the unsupported region by an external stimulus, for example, the electric field of metallic gates \cite{Chen2009, Wong2010, Xu2010, Lee2013, Mathew2016, Alba2016, Davidovikj2016, Davidovikj2018, Guettinger2017}, by pressurized air \cite{Shin2016,  Smith2016} or by periodic driving at the membrane eigenfrequencies \cite{Atalaya2008, Davidovikj2016}, and forms a nanoelectromechanical system. Mechanical driving leads to membrane oscillations which, due to the ultra-high electron mobility in graphene, can be treated as adiabatic changes of the background geometry in which the electronic current adjusts immediately to the quasi-static deformations \cite{Eriksson2013, Zhang2015, Popescu2016}. Optical driving of the graphene membrane is also possible but this leads in general to very small deformations \cite{Miller2019}. The membranes can be sealed (or clamped) \cite{Chen2013, Lee2013, Lee2019}, and therefore, are also called graphene nanodrums \cite{Davidovikj2016, Davidovikj2018}. The deformation of the graphene membrane generates strain in the material. Its gradient induces a strong pseudomagnetic field which acts with opposite signs on the electrons in the two valleys \cite{Vozmediano2010, Stegmann2016, Naumis2017}, and thus, can be used to spatially separate valley polarized currents.

The appearance of strong pseudomagnetic fields in deformed graphene has been previously utilized to propose valley filters \cite{Wang2014, Settnes2016, Stegmann2019, Milovanovic2016, Carrillo-Bastos2016, Zhai2018, Carrillo-Bastos2018, Mahmud2020, Solomon2021} but to the best of our knowledge these devices have not yet been realized experimentally. One of the main obstacles is the challenge to fabricate properly deformed or strained graphene sheets, despite certain recent progress \cite{Dai2019}. We demonstrate that graphene nanoresonators with already low strain values of the order of $0.1 \sim 1\%$ can be used to construct efficient valley polarizers. Even lower strain values can be used due to the possibility to place various resonators in series \cite{Zande2010,  Barton2011, Liu2013}. Moreover, we show that the proposed devices can be applied to collimate and direct electrons and thus control the electronic transport.

\section{Model of the graphene membrane}

We consider graphene membranes with a size of about $300 \un{nm} \times 600 \un{nm}$, which are deposited on an insulating substrate with a cavity, see \fig{1}. We assume that the membrane is clamped at the edges of the cavity, for example, by means of photoresist \cite{Chen2013, Lee2013, Lee2019}. The graphene membrane is modelled by the typical first nearest neighbor tight-binding Hamiltonian
\begin{equation}
  \label{eq:1}
  H= -t_0 \sum_{i,j} \ket{i^A}\bra{j^B} +\text{H.c.}
\end{equation}
The $\ket{i^{A/B}}$ indicate the atomic states localized on the carbon atoms at positions $\v r_i$ in the sub-lattices A and B, respectively. The sum runs over nearest neighboring atoms, which are separated by the distance of $d_0=0.142 \un{nm}$ and coupled with the energy $t_0= 2.8 \un{eV}$. The minor effects of the substrate on the graphene membrane are neglected.

First, we consider circular cavities of radius $r_0$ where the deformation of the graphene membrane is given by the circular drum modes
\begin{equation}
  \label{eq:2}
  h_{mn}(r,\phi)= 
  \begin{cases}
    a \, J_m(\lambda_{mn} r/r_0)\cos(m \phi) &\text{if} \: r\leq r_0,\\
    0 \quad &\text{otherwise,}
  \end{cases}
\end{equation}
where $J_m$ is the $m$-th Bessel function of the first kind and $\lambda_{mn} $ is the $n$-th zero of $J_m$. The amplitude of the deformation is controlled by the parameter $a$. Second, we consider also rectangular shaped cavities of size $\mathcal{L}= L_x \times L_y$ with the deformation given by the rectangular drum modes
\begin{equation}
  \label{eq:3}
  h_{mn}(x,y)=
  \begin{cases}
  a\, \sin(n\, x/L_x) \sin(m\, y/L_y) &\text{if} \: x,y \in \mathcal{L},\\
    0 \quad &\text{otherwise.}
  \end{cases}
\end{equation}
The deformations enlarge the distance of neighboring carbon atoms and weaken their coupling. This modification is in good approximation described by
\begin{equation}
  \label{eq:3}
  t_{ij} \cong t_0 \exp(-\bt\, \dl_{ij}),
\end{equation}
where $\dl_{ij}= \frac{\abs{\v r_i- \v r_j}-d_0}{d_0}$ and $\bt=3.37$ \cite{Pereira2009,   Ribeiro2009, Carrillo-Bastos2016}. These deformations are motivated by the classical membrane modes and could be generated by mechanical periodic driving of the system at the corresponding eigenfrequencies \cite{Atalaya2008, Davidovikj2016} or by pressurized air \cite{Shin2016, Smith2016}.

In both cases, the real atomic structure of graphene can slightly deviate from our model, in which we assume a simple lift of atoms according to $z=h(x,y)$ and creating strong strain at the boundary where the slope of $h$ is the largest. A structural relaxation process using molecular dynamics may be applied to the deformed membrane in order to get closer to the experimental situation. We plan to address this point in our future work.

\section{Electronic transport in graphene}
\subsection{The Green's function method}

The current flow in graphene NEMS is studied by means of the Green's function method. Here we summarize the essential equations of this method as detailed introductions can be found in various textbooks \cite{Datta1997, Datta2005}.

The Green's function of the system is given by 
\begin{equation}
  \label{eq:9}
  G(E)= \lr{E-H-\Sg}^{-1},
\end{equation}
where $E$ is the energy of the injected electrons and $H$ is the tight-binding Hamiltonian, \eq{1}. The self-energy $\Sg= -\I \sum_{i \in \text{edges}} \ket{i}\bra{i}$ is a complex potential at the edges of the system which absorbs the electrons and suppresses finite-size
effects.

Electrons are injected and detected through contacts at the edges of the system, see the golden bars in \fig{1}. We use two different models for the injection of the electrons. In the first one, the electrons are injected as plane waves represented by the inscattering function
\begin{equation}
  \label{eq:10}
  \Sg^{\text{in}}_{\text{pw}}= \sum_{i,j \in \text{contact}} A(\v r_i) A(\v r_j)
  \psi_{j}^*(\v k) \psi_{i}(\v k) \ket{i} \bra{j},
\end{equation}
where the sum runs over the carbon atoms at the contact, see \fig{1}. The $\psi_i(\v k)$ are the plane-wave eigenstates of graphene's Dirac Hamiltonian (see \eq{18-x} below)
\begin{equation}
  \label{eq:11}
  \psi_i(\v k){=}
  \begin{cases}
    c^{-} e^{\I(\v k{+}\v K^-)\v r_i} + c^{+} e^{\I(\v k{+}\v K^+)\v r_i} & i \in A,\\
    s\, c^{-} e^{\I(\v k{+}\v K^{-}) \v r_i + \I \phi} \,{-}\, s\, c^{+}
    e^{\I(\v k{+}\v K^{+}) \v r_i {-} i\phi} \hspace*{-2mm}& i \in B,
  \end{cases}
\end{equation}
where $\phi= \arg(\I k_x + k_y)$. The parameters $c^\pm$ control the occupation of the two $\v K^\pm$ valleys. In the following, we consider that for the injected electrons both valleys are fully mixed, i.e.  $c_\pm= \pm 1/2$. The function
\begin{equation}
  \label{eq:12}
  A(\v r)= e^{-\lr{\v k \cdot (\v r- \v r_0)/w_0}^2}
\end{equation}
gives the injected current beam a Gaussian profile. The parameters $\v r_0$ and $w_0$ control the position and width of the beam. The advantage of this model is that it allows to inject narrow electron beams with given energy, momentum and valley polarization. Such beams are ideal to compare the current flow with semiclassical trajectories and phenomena from optics \cite{Stegmann2016, Stegmann2019, Betancur2019, Betancur2020, Paredes2021}.

In the second case, we will use the so-called wideband model
\begin{equation}
  \label{eq:12b}
  \Sg^{\text{in}}_{\text{wb}}= \sum_{i,j \in \text{contact}} \eta\ket{i} \bra{j},
\end{equation}
where the injecting contact is characterized by a constant, energy independent surface density of states, $\text{DOS} \propto \eta = \text{const}$. Through this contact unpolarized electrons of energy $E$ are injected without a precisely specified momentum leading to a strongly divergent electron beam. This generic model represents well experimental situations where it is not precisely known how the electrons are entering the nanosystem via the contacts.

Finally, the current flowing between the atoms at positions $\v r_i$ and $ \v r_j$ is calculated by
\begin{equation}
  \label{eq:13}
  I_{ij} = \textrm{Im}(t_{ij}\, (G\, \Sigma^{\text{in}}\, G^{\dagger})_{ij}).
\end{equation}
This bond current is then averaged (or coarse grained) over the six edges of the carbon atoms.

\subsection{Valley polarization}

The valley polarization of a state $\ket{\phi}$ characterizes to which degree this state occupies the two valleys $\v K^\pm$ in graphene. It can be calculated by the projection $\mc{P}(\v k)= \abs{\braket{\psi(\v k)| \phi}}^2$ onto the graphene eigenfunctions \eq{11}. Within the Green's function approach this projection reads \cite{Stegmann2019, Settnes2016}
\begin{equation}
    \label{P_i_k}
    P_{i}(\v k)=\braket{\psi(\v{k})|G\, \Sigma^{\text{in}}\, G^{\dagger}|\psi(\v{k})}_{A_i}.
\end{equation}
It is calculated over a finite region $A_i$ of the system, see for example the gray-shaded regions in \fig{4}, in order to assess the valley polarization locally. The spectral density $P_i(\v k)$ is integrated around the valleys
\begin{equation}
    \label{P_+-}
    \mc{P}_{i}^{\pm}=\int_{\v{k}\in \v{K}^{\pm}}d^2k \, P_{i}(\v{k})
\end{equation}
and the valley polarization is given by
\begin{equation}
    \label{Pi}
    \mc{P}_{i}=\frac{\mc P_{i}^{+}-\mc P_{i}^{-}}{\mc P_{i}^{+}+\mc P_{i}^{-}}.
\end{equation}
For $\mc P_i = \pm 1$ the electrons are localized exclusively at the $\v{K}^\pm$ valleys and hence, are completely valley polarized, whereas for $\mc P_i=0$ they are completely unpolarized.

\section{Continuous model of graphene membrane}
\subsection{Dirac equation for deformed graphene}

At low energies, where the electron wavelength is larger than the lattice constant the discrete tight-binding Hamiltonian \eq{1} can be approximated by the continuous Dirac Hamiltonian \cite{Juan2012, Juan2013, Oliva-Leyva2015, Stegmann2016}
\begin{equation}
  \label{eq:18-x}
  H^D = \I \hbar v_F \sg^a e_a^{\;\;l}(\v x) \lr{\pd_l - \I K^{\pm}_l(\v x)}
\end{equation}
describing relativistic massless fermions. Here, $v_F=3 t_0 d_0/2\hbar$ is the Fermi velocity of the electrons and $\sg^a$ ($a=1,2$) are the Pauli matrices. The local frame vectors
\begin{equation}
  \label{eq:local-frame}
  \v e_a(\v x) = \lr{1 - \bt \hat\eps(\v x)} \v e_a,
\end{equation}
describe the curvature of the membrane and are determined by the strain tensor 
\begin{equation}
    \label{eq:strain}
    \eps_{ij} = \frac{1}{2}\, \pd_i h(\v x) \pd_j h(\v x)    
\end{equation}
multiplied by the factor $\beta>1$. Hence, the effective deformation for the electrons is magnified by $\beta$ but otherwise identical to the real geometry of the deformed graphene membrane.

Due to the deformation, the two valleys, located in pristine graphene at $\v K^\pm= (0, \pm\frac{4 \pi}{3\sqrt{3} d_0})$, become a function of the position inside the device \cite{CastroNeto2009, Vozmediano2010}
\begin{equation}
  \v K^\pm(\v x) = \v K^\pm \pm \frac{\beta}{2} 
  \bigl(-2 \eps_{xy}, \eps_{yy}-\eps_{xx}\bigr).
\end{equation}
This function can be interpreted formally as a vector potential and its curl causes an effective pseudomagnetic field
\begin{equation}
  \label{eq:Bfield}
  B^\pm(\v x) =  \pm \frac{\beta}{2}\bigl(\pd_x\, \eps_{yy}(\v x) - \pd_x\, \eps_{xx}(\v x)  + 2\pd_y \eps_{xy}(\v x)\bigr),
\end{equation}
which is perpendicular to the graphene plane. In contrast to a true magnetic field, the pseudomagnetic field acts with opposite sign in the two different valleys, preserving the time-reversal symmetry of the system. This sign change of the pseudo magnetic field will be used to separate spatially the electrons from different valleys.

\subsection{Current flow lines in geometric optics approximation}

For deformations larger than the electron wavelength the geometric optics approximations can be applied to the Hamiltonian \eq{18-x}. In our previous work \cite{Stegmann2016}, we have shown that in this case the current flow can be predicted by the semi-classical trajectories of relativistic massless fermions
\begin{equation} 
  \label{eq:geodesic}
  \frac{dv^i}{d\tau} = - \Gamma^i_{kl} v^k v^l + \sqrt{g}\ g^{ij} \epsilon_{jk}\, v^k B^{\pm}
\end{equation}
where $v^i(\tau) = dx^i(\tau)/d\tau$ is the ``velocity''. The fist term on the right-hand side takes into account the curvature through the Christoffel symbols
\begin{equation}
    \Gamma_{kl}^i= \frac{1}{2} g^{ij} \lr{\pd_k g_{jl} + \pd_l g_{kj} - \pd_j g_{kl}}.
\end{equation}
The second term describes the electromagnetic force where the $g^{ij}= \delta_{ij} -2 \beta \eps_{ij}(\v x)$ are the effective (inverse) metric and $\epsilon_{ij}$ is the Levi-Cevita symbol in two dimensions. The calculation of these trajectories is computationally much less demanding than the quantitative quantum approach and independent from the system size. Therefore, it provides a useful tool to estimate the current flow in deformed graphene nanostructures.

\begin{figure}[t]
  \centering
  \includegraphics[width=0.95\linewidth]{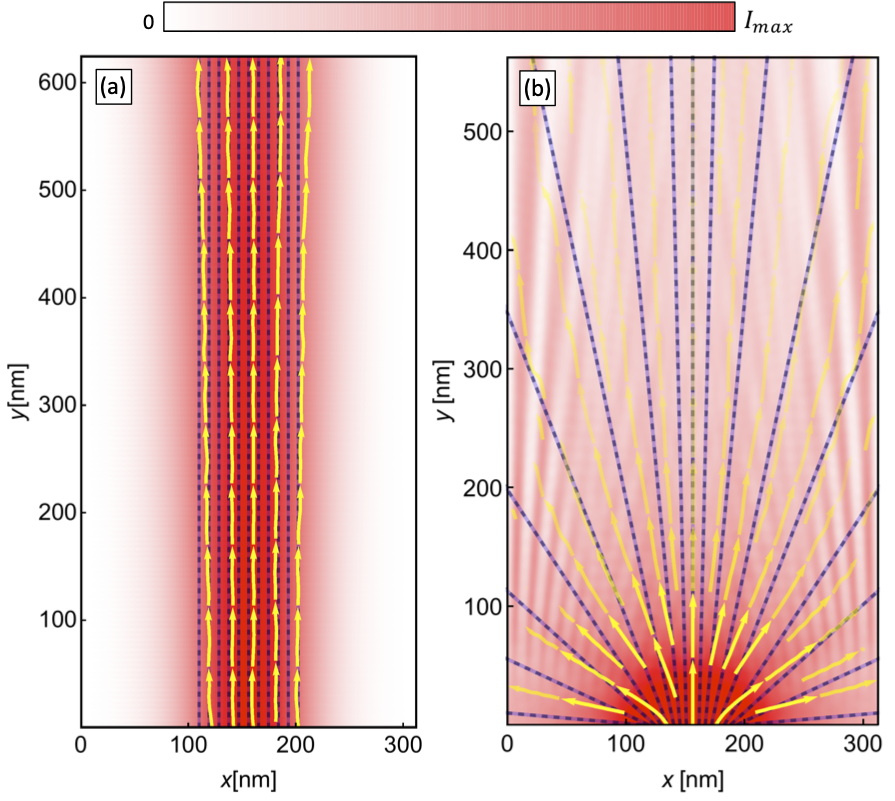}
  \caption{Current flow in a pristine graphene membrane. The electrons are injected at the bottom armchair edge. The current vector field is given by the yellow arrows, its norm by the red color shading. Solid blue lines and dashed black lines are semiclassical trajectories for electrons in valley $\v K^+$ and $\v K^-$, respectively. (a) Ballistic beam-like current propagation is observed for electrons injected by the plane-wave model at energy $E=200\un{meV}$. (b) In the case of the generic wide-band model the electrons, injected at a slightly lower energy of $E=170\un{meV}$, are dispersed strongly.}
  \label{fig:2}
\end{figure}

\begin{figure*}[t]
  \centering
  \includegraphics[scale=0.3]{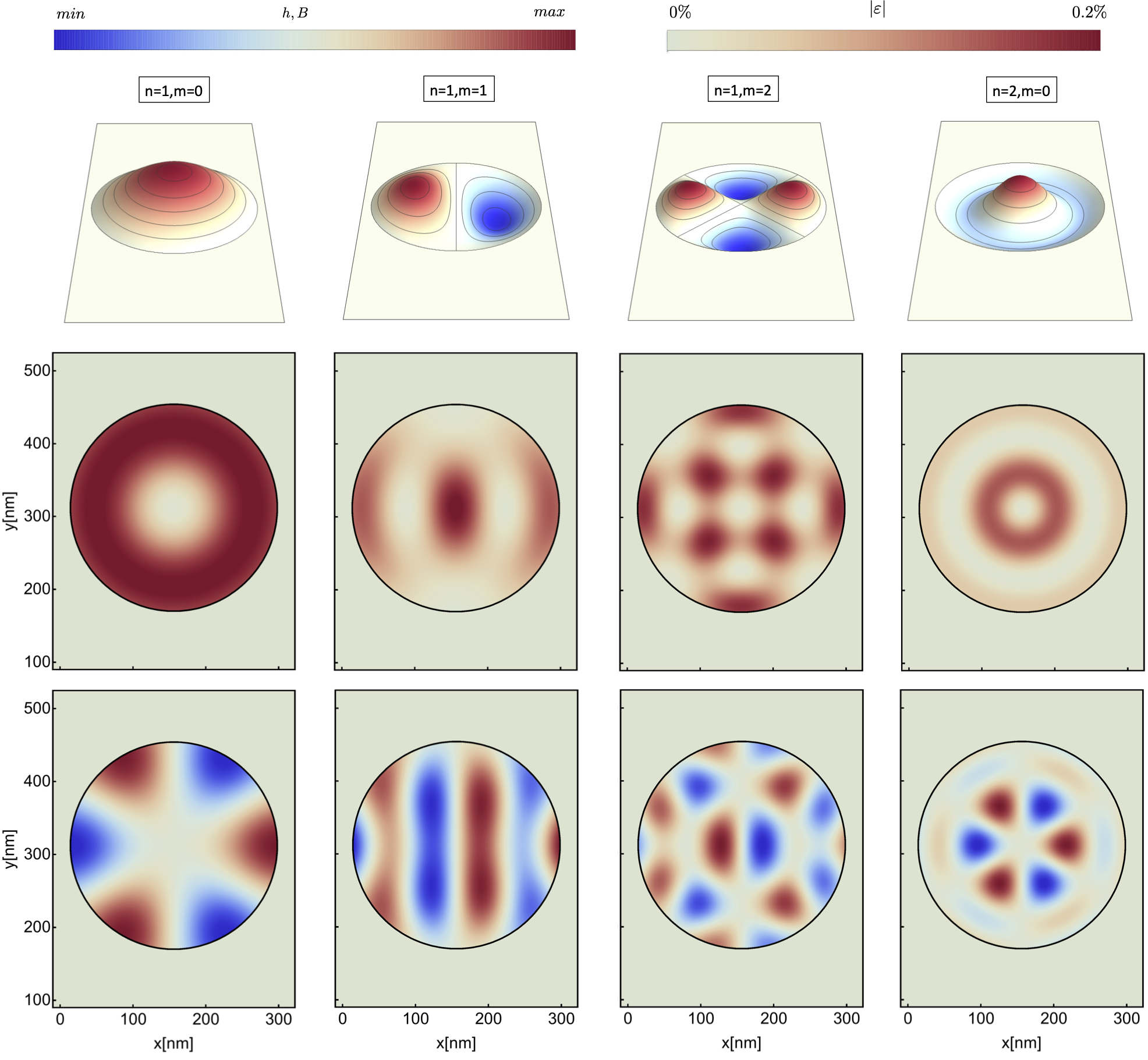}
  \caption{Different modes $(n,m)$ of a circular graphene nanodrum. The deformation profiles are shown in the top row. The resulting strain and pseudomagnetic fields can be found in the middle and bottom row, respectively. The different drum modes lead to very distinct pseudomagnetic field patterns.}
  \label{fig:3}
\end{figure*}

\section{Results}

We begin our discussion with \fig{2} showing the current flow in a pristine sheet of graphene. The electrons are injected at the bottom edge which has the armchair shape. The current vector field is visualized by the yellow arrows and its norm, the current density, by the red color shading.\footnote{As the current shows some rare peaks, its maximum value is defined here as four standard deviations above the mean.} On the left-hand side, the electrons are injected by means of the plane-wave model resulting in a ballistic beam-like propagation through the device. On the right-hand side, the generic wide-band model is used for the injection of the current which leads to a much more dispersed current flow. In this case, the electrons are partially reflected at the lateral edges (despite of the absorbing contacts) leading to a ripple pattern in the current density due to interference. The current flow in graphene membranes will be described by one of the two models, depending on the experimental realization of the contacts.

We study here only static configurations since any graphene membrane dynamics (with oscillation frequencies ranging between $10$ to $100 \un{MHz}$ for probes of size of $\sim 1 \un{\mu m}$ \cite{Cooper2012, Shi2019}) is so much slower than the time-scale dictated by the electron mobility ($\sim 1$ THz) that it can be treated as adiabatic change of the geometry to which the electronic current adjusts immediately \cite{Eriksson2013, Zhang2015, Popescu2016}.

\begin{figure*}[t]
  \centering
  \includegraphics[width=\textwidth]{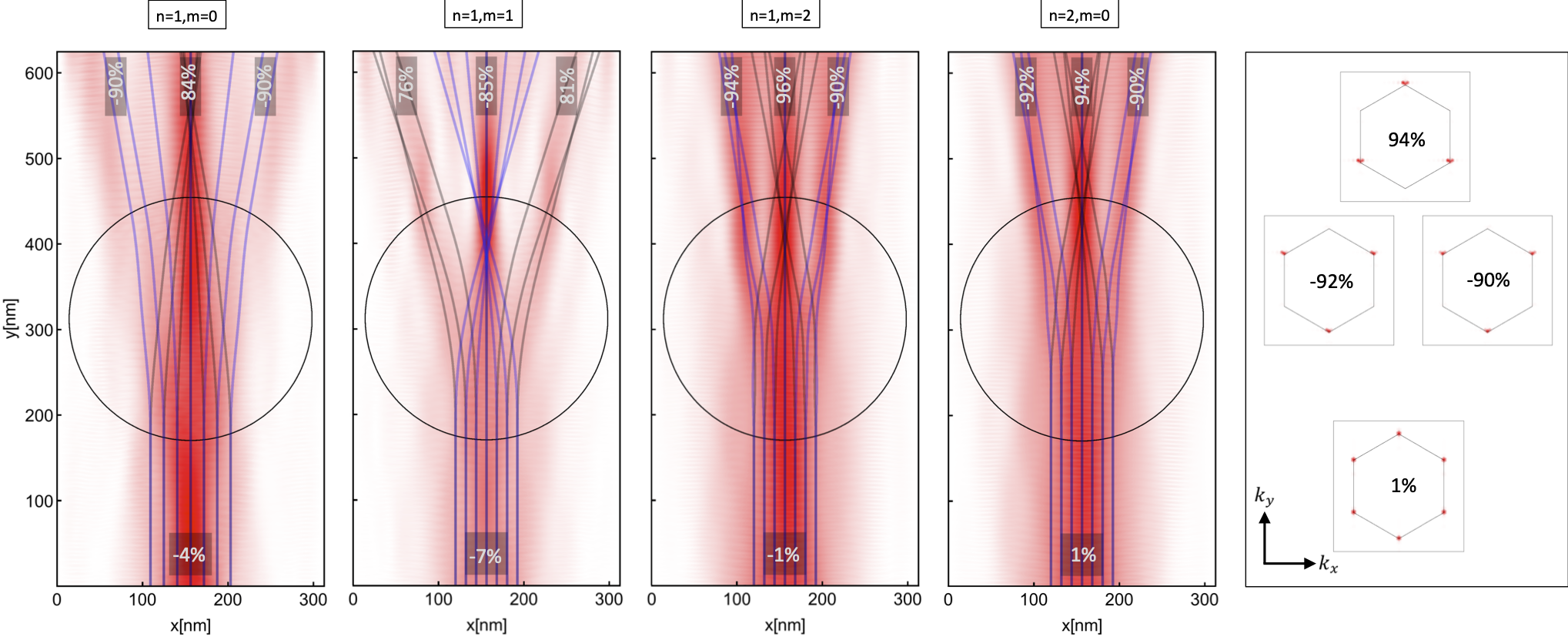}
  \caption{Current flow in different drum modes $(n,m)$ of circular graphene nanoresonators with a maximum strain ranging between $0.13$ and $0.2\%$. The electrons, injected at $E=200\un{meV}$ by the plane-wave model, are split into three beams. The black and blue solid lines are semi-classical trajectories for electrons in the $\v K^+$ and $\v K^-$ valley, respectively. These trajectories agree qualitatively with the current flow patterns and can be used to estimate the electronic transport in the device. The spectral densities $P_i$ are calculated within the gray shaded rectangles and shown for the $(2,0)$ mode at the right. The resulting polarizations are given in percentages within the gray shaded rectangles and show that an initially unpolarized current is split in three valley polarized beams.}
  \label{fig:4}
\end{figure*}

\subsection{Circular graphene nanodrums}

\begin{figure*}[t]
  \centering
  \includegraphics[width=\textwidth]{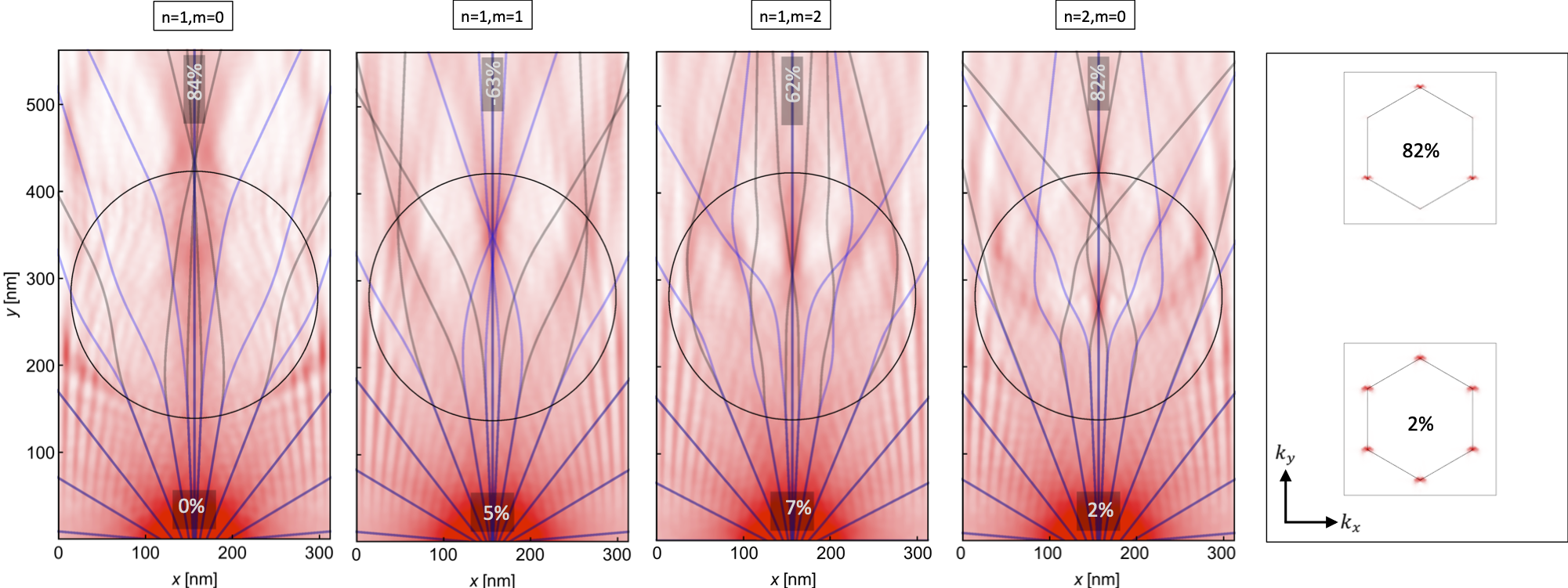}
  \caption{Current flow for electrons injected at $E=170\un{meV}$ by the wide-band model. The maximal strain is in between $0.53$ and $0.66\%$. The current is much more dispersed compared to the plane wave model but part of the current is focused onto a narrow valley polarized beam that is passing through the center of the nanodrum. The current flow pattern agree qualitatively with the semiclassical trajectories of strongly dispersive electrons. The spectral densities within the gray shaded regions of the $(2,0)$ mode are shown at the right. The resulting polarizations are indicated in percentages.}
  \label{fig:5}
\end{figure*}

We proceed with membranes deposited on a circular cavity forming a GrNEMS or graphene nanodrum. \Fig{3} shows the deformation profile (top row), the strain (middle row) and the pseudomagnetic field (bottom row) for various drum modes $(n,m)$. The figures demonstrate that the shape of the pseudomagnetic field, which is perpendicular to the graphene sheet, changes drastically with the different drum modes. 

\begin{figure*}[t]
  \centering
  \includegraphics[scale=0.7]{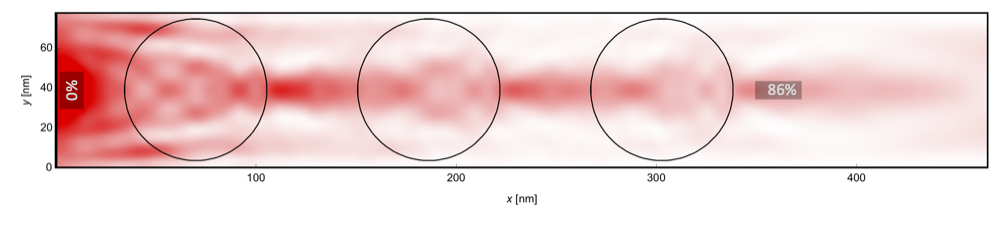}
  \caption{Current flow in a device of three graphene nanodrums in series. The membrane is in the (1,0) mode with a maximum strain of $0.63\%$. The strongly dispersed electron beam, injected at the left edge, is strongly collimated by the resonator demonstrating the rich functionality of nanodrum devices.}
  \label{fig:6}
\end{figure*}

\begin{figure*}[t]
  \centering
  \includegraphics[scale=0.2]{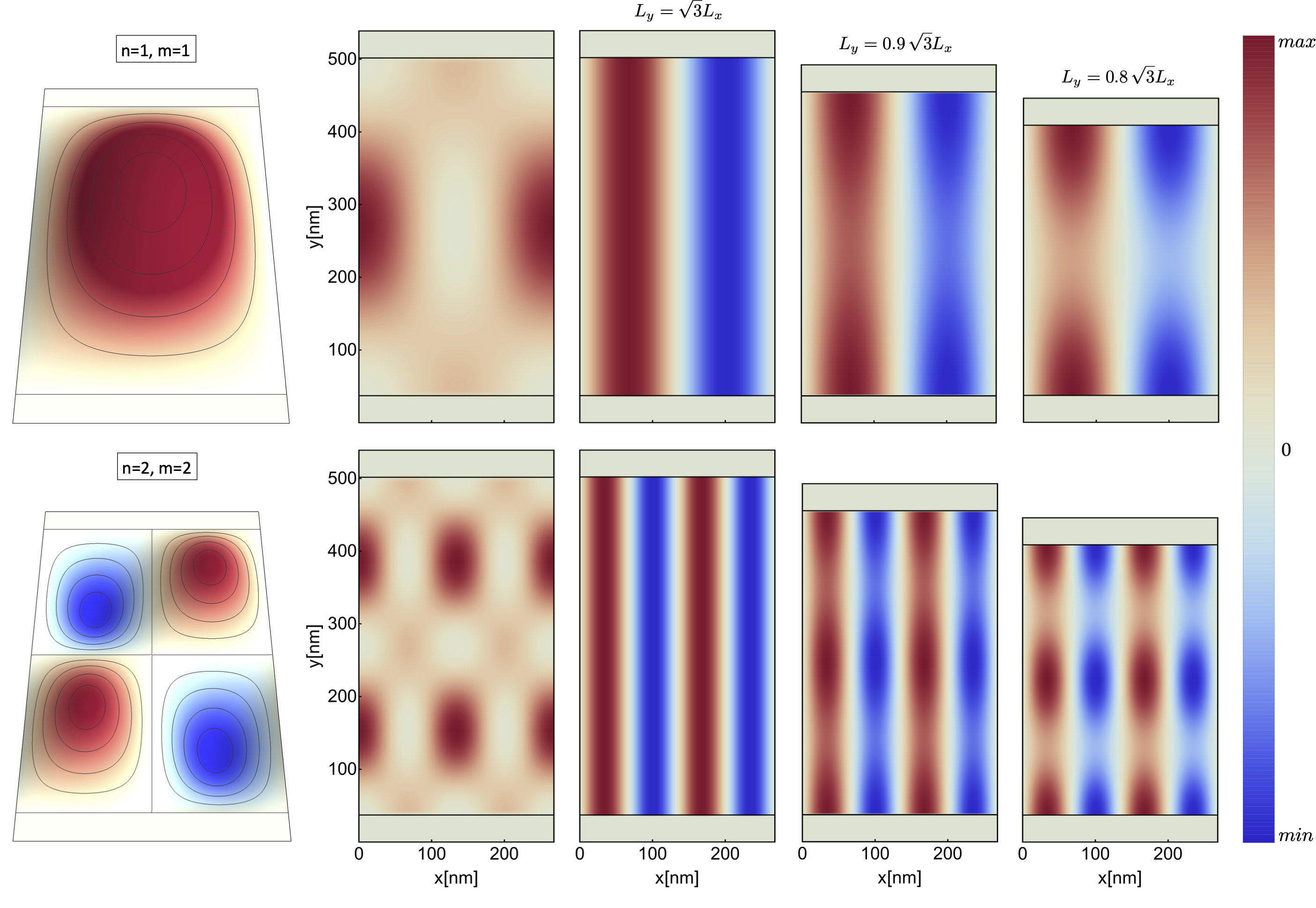}
  \caption{Deformation profile (first column), strain (second column) and pseudomagentic field (third column) of the $(n,m)= (1,1)$ and $(2,2)$ mode of a rectangular graphene nanoresonator. For an aspect ratio of $L_y/L_x= \sqrt{3}\,m/n$ the pseudomagnetic field is independent from the $y$ coordinate and has a stripe-like shape. The alternating sign of the pseudomagnetic field in the strips is optimal to confine valley polarized electron beams. The shape of the pseudomagnetic field is rather robust against variations of the aspect ratio, compare the third, fourth and fifth column. 
  }
  \label{fig:7}
\end{figure*}

The corresponding current flow pattern for electrons injected by the plane-wave model at energy $E=200 \un{meV}$ are shown in \fig{4}. It can be observed that the ballistic current is separated in three beams; two are deflected to the edges, while one goes through the center of the drum. The spectral density $P_i(\v k)$ of these current beams is calculated in the gray shaded rectangular regions and shown in \fig{4} (right) for the $(2,0)$ mode. Close to the injecting source contact at the bottom edge, we observe small red dots in all six edges of graphene's Brillouin zone, which indicate that the electrons occupy both valleys and the current is unpolarized. In the regions at the upper edge, we observe red dots only at three equivalent Dirac points, proving that the current is fully valley polarized. Note that the dots in the central and in the outer regions are located in different valleys and thus, the current in these regions is polarized in the opposite way. The spectral density, integrated around the different valleys, leads to the polarization $\mc{P}_i$, which is given in percentages in the rectangular regions of the current flow patterns. In general, we find that an initially unpolarized current is converted to a highly valley polarized current $\mc{P}_i> 85\%$, which demonstrates clearly that the proposed device can be used as a valley polarizer or valley filter. The strain of the graphene membrane in the different drum modes lies in the range of $0.13$ to $0.2\%$. This is an important feature of the proposed device because the experimentally achievable strain values are limited to some few percents or to values even below, depending on the used setup \cite{Shin2016,  Smith2016}.

The semi-classical trajectories, which are given by the solution of the differential equation \eq{geodesic}, are indicated in \fig{4} by the solid blue and black curves for electrons from the $\v K^+$ and $\v K^-$ valleys, respectively. These trajectories follow qualitatively the quantum current densities. Their behavior can be understood largely by taking into account the Lorenz force due to the pseudomagnetic field (see \fig{3}) acting on the electrons of different valleys with opposite signs. This Lorenz force focuses the electrons from one valley on a narrow beam that passes straight though the resonator, while the electrons from the other valley are deflected towards the edges. The trajectories also allow to understand qualitatively the observed valley polarizations. Note that in the $(1,1)$ mode the sign of the polarization is reversed compared to the other modes.

In the case of electron injection by the wide-band model, see \fig{5}, the current flow is much more dispersed as in this model the electron energy is fixed ($E=170 \un{meV}$) but the momentum vector is unspecified. Nevertheless, we observe that part of the current density is deviated towards the edges, while another fraction of the current is focused on a beam that goes through the center of the resonator. The current flow can be understood qualitatively by the semiclassical trajectories if the initial spreading of the injected electron beam is taken into account. The valley polarization calculated at the bottom and top of the device confirms that valley polarization of up to $84\%$ can be obtained, which is a surprisingly high value considering that the electrons are strongly dispersed and partially reflected at the system edges. The used strain values, which range between $0.53$ and $0.66\%$, are slightly higher, because stronger pseudomagnetic field are required to focus the dispersive electron beam. The functionality of the device can be further improved by placing several nanoresonators in series, generating a valley polarized and collimated current beam, see \fig{6}. Note that in \fig{6} the device is rotated by $90^\circ$ but the current is still injected at the armchair edge.

\subsection{Rectangular graphene nanoresonators}

\begin{figure}[t]
  \centering
  \includegraphics[scale=0.25]{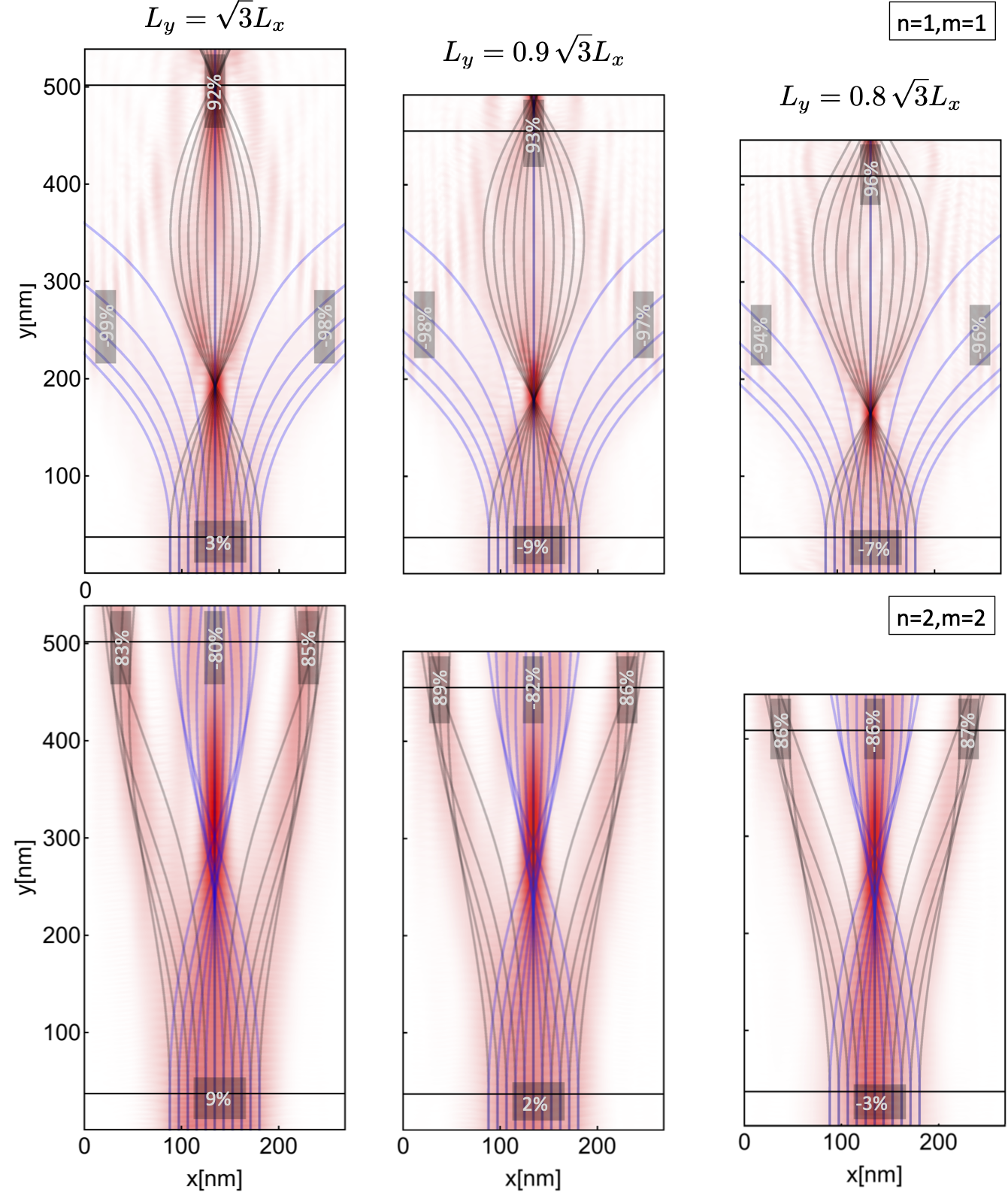}
  \caption{Current flow in the $(n,m)= (1,1)$ and $(2,2)$ modes of a rectangular graphene nanoresonator. Electrons are injected at energy $200\un{meV}$ by the plane-wave model. The edges of the resonator are indicated by black horizontal lines and the vertical system edges. The valley polarization is calculated and indicated within the gray shaded rectangular regions. The generation of narrow valley polarized electron beams is observed which are stable if the aspect ratio is changed from its optimal value $L_y/Lx=\sqrt{3}\,m/n$. The semiclassical trajectories, indicated for the electrons of different valley by the solid blue and black curves, follow qualitatively the current density. In particular, their crossing points agree with the focusing points of high current density.}
  \label{fig:8}
\end{figure}

We continue our study with rectangular graphene nanoresonators. The deformation, strain and pseudomagentic field patterns are shown in \fig{7} for two different drum modes. Most notable is the fact that for an aspect ratio of $L_y/L_x= \sqrt{3}\,m/n$ the pseudomagnetic field is constant in the $y$ direction. This property, which can be proven easily by using Eqs. \eq{3}, \eq{strain} and \eq{Bfield}, is rather stable and only lost slowly when the aspect ratio changes. 

The current flow patterns for the plane-wave model of injection are shown in \fig{8}. In the case of the $(1,1)$ mode, where a maximum strain of $1.5\%$ is present, the electrons from one valley are focused on a narrow beam in the center, while the electrons form the other valley are deviated towards the edges of the system. Note that these electrons are partially reflected at the edges indicating that the absorption of the complex potential is not perfect. In the case of the $(2,2)$ mode a maximum strain of $0.24\%$ is applied and the electrons are split into three narrow beams. The semiclassical trajectories agree well with the current flow. Most notably, the crossing points of the trajectories agree with the focusing points of the current density. The valley polarizations, measured and indicated in the gray rectangular regions in \fig{8} demonstrate that highly valley polarized electron beams are generated with polarizations above $80\%$. Moreover, the current flow patterns and valley polarizations persist when the aspect ratio is changed from its optimal value. The strip-like pseudomagnetic fields favor the generation of narrow beams of valley polarized electrons. In the case of the $(1,1)$ mode electrons from one valley are sorted out towards the edges, while in the $(2,2)$ three valley polarized beams are obtained. 

\begin{figure}[t]
  \centering
  \includegraphics[scale=0.25]{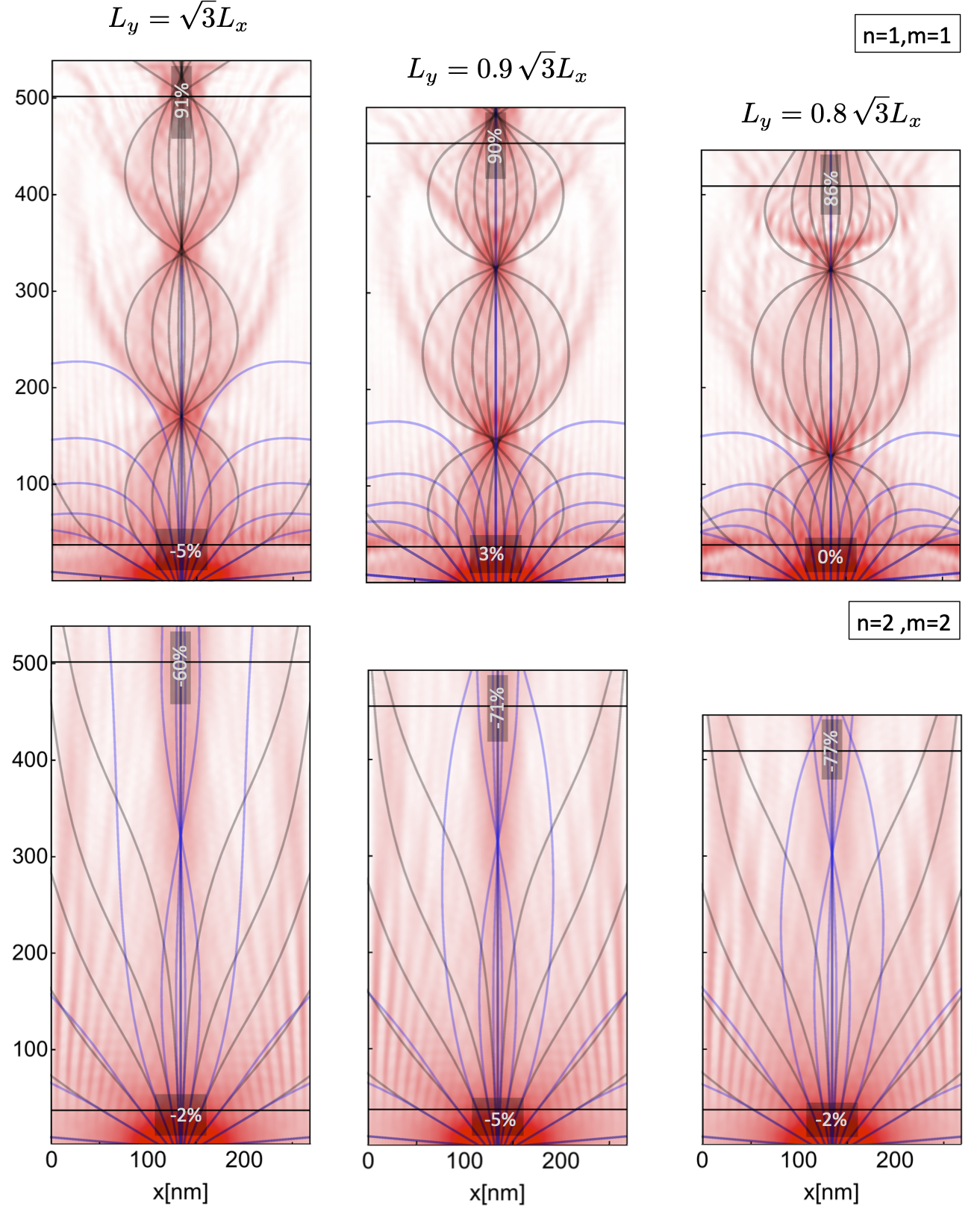}
  \caption{Current flow for electrons injected by the wide-band model at $E=170\un{meV}$. Part of the current density is confined in the central region of the device. Valley polarization of about $90\%$ in the $(1,1)$ mode and $70\%$ in the $(2,2)$ mode are found. Due to the spreading of the electron beam stronger pseudomagnetic fields are necessary and in consequence, higher strain values of $4.4\%$ and $0.35\%$, respectively. The semiclassical trajectories follow qualitatively the current density. Most impressively, their consecutive crossing points match with the focusing points of the current density.}
  \label{fig:9}
\end{figure}

When we proceed with the wide-band model of electron injection, see \fig{9}, the current becomes more dispersed (as in the case of circular nanodrums) but part of it is confined in a narrow region in the center of the device. This current shows valley polarizations of about $90$ \% and $70 \%$ in the $(1,1)$ and $(2,2)$ modes, respectively. However, this goes along with higher strain values of $4.4\%$ and $0.35\%$ for the two modes, because stronger pseudomagnetic fields are necessary to focus the dispersed current flow and valley polarize the electrons. The semiclassical trajectories agree well with the current density. In particular, the consecutive crossing points of the trajectories match precisely with the focusing points of the current density.

\begin{figure}[t]
  \centering
  \includegraphics[scale=0.4]{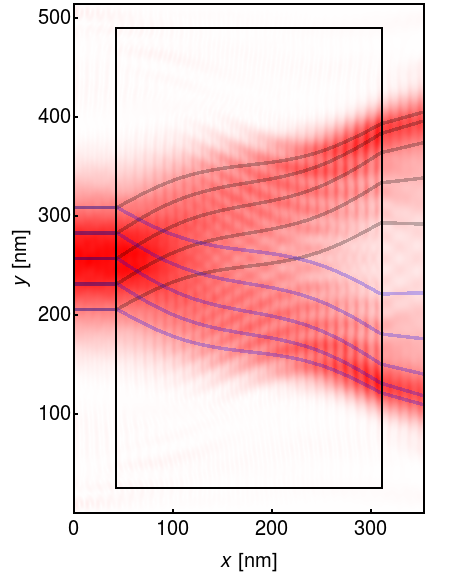}
  \caption{Current flow for electrons injected at the left zigzag edge at $E=200\un{meV}$ using the plane-wave model. The abrupt change of the strain at the left edge of the resonator (cf. \fig{7}) generates formally a singular pseudomagnetic field (along the black line) and leads to a sudden current splitting for which the semiclassical trajectories must be specially adjusted.}
  \label{fig:10}
\end{figure}

\subsection{Current injection at zigzag edge}

In all devices presented in the previous sections the current has been injected at the armchair (bottom) edge. It might be tempting to inject the current also at the zigzag (left) edge and expect a clear separation of the valley polarized beams. Due to the even number of stripes with non-vanishing pseudomagnetic field $B^\pm(\v x)$ the beams should finally become almost parallel, see the semiclassical trajectories in \fig{10}.

However, the situation at the zigzag boundary is very different from the armchair case. The pseudomagnetic vector potential $\v K^\pm(\v x)$ is always parallel to the zigzag direction (i.e. has only a $y$-component) and has a jump at the zigzag boundary which leads to formally infinite pseudomagnetic field $B^\pm(\v x)$. This forces the plane wave to abruptly change its propagation direction, depending on the valley polarization. At the same time, the strain tensor $\epsilon_{ij}(\v x)$ has a jump, too. Since these effects happen on a small scale, the continuous approximation can break down and the current flow can differ significantly from the semiclassical trajectories. Nevertheless, after adjusting the trajectories to the kick-like velocity change, we find a quite good agreement between the quantum and semiclassical currents (see \fig{10}). The abrupt change of the strain is an artefact of our simple model of the graphene nanodrum. In order to obtain a more realistic picture one should consider smoother edges of the membrane. The above mentioned relaxation of the atomic structure should modify the strain distribution exactly in this direction. This point is in focus of our ongoing research and will be addressed elsewhere.

\section{Conclusions}

We have demonstrated that GrNEMS can be used as efficient valley polarizers. In the case of circular graphene nanodrums strain values already between $0.1$ and $0.6\%$ lead to highly polarized currents. The device works for the plane-wave model, where the electrons are injected as a narrow ballistic beam, as well as for the wideband model, where the injected current is strongly dispersive. In the latter case, higher strain values and thus stronger pseudomagnetic fields are able to
not only valley polarize the electrons but also focus them to a narrow beam acting as a beam collimator. In the case of rectangular nanoresonators, we found that a special aspect ratio of $L_y/L_x=\sqrt{3}\, m/n$ for the drum mode $(m,n)$ leads to a stripe-like pseudomagnetic field, constant along the $y$ direction, which optimally confines the valley polarized electrons beams. The semiclassical trajectories agree well with the current calculated by means of the Green's function method. Most impressively, their crossing points agree precisely with the focusing points of the current density. These trajectories represent a computationally efficient tool to predict the current flow in graphene nanostructures. 

\section{Acknowledgments}

WO gratefully acknowledges a CONACYT graduate scholarship. WO and TS gratefully acknowledge funding from CONACYT Proyecto A1-S-13469, and UNAM-PAPIIT IN103922. NS gratefully acknowledges funding by the Deutsche Forschungs\-gemeinschaft (DFG, German Research Foundation) -- Project 278162697 -- SFB 1242.
  
\bibliography{nmrvp}

\begin{thebibliography}{54}%
\makeatletter
\providecommand \@ifxundefined [1]{%
 \@ifx{#1\undefined}
}%
\providecommand \@ifnum [1]{%
 \ifnum #1\expandafter \@firstoftwo
 \else \expandafter \@secondoftwo
 \fi
}%
\providecommand \@ifx [1]{%
 \ifx #1\expandafter \@firstoftwo
 \else \expandafter \@secondoftwo
 \fi
}%
\providecommand \natexlab [1]{#1}%
\providecommand \enquote  [1]{``#1''}%
\providecommand \bibnamefont  [1]{#1}%
\providecommand \bibfnamefont [1]{#1}%
\providecommand \citenamefont [1]{#1}%
\providecommand \href@noop [0]{\@secondoftwo}%
\providecommand \href [0]{\begingroup \@sanitize@url \@href}%
\providecommand \@href[1]{\@@startlink{#1}\@@href}%
\providecommand \@@href[1]{\endgroup#1\@@endlink}%
\providecommand \@sanitize@url [0]{\catcode `\\12\catcode `\$12\catcode
  `\&12\catcode `\#12\catcode `\^12\catcode `\_12\catcode `\%12\relax}%
\providecommand \@@startlink[1]{}%
\providecommand \@@endlink[0]{}%
\providecommand \url  [0]{\begingroup\@sanitize@url \@url }%
\providecommand \@url [1]{\endgroup\@href {#1}{\urlprefix }}%
\providecommand \urlprefix  [0]{URL }%
\providecommand \Eprint [0]{\href }%
\providecommand \doibase [0]{https://doi.org/}%
\providecommand \selectlanguage [0]{\@gobble}%
\providecommand \bibinfo  [0]{\@secondoftwo}%
\providecommand \bibfield  [0]{\@secondoftwo}%
\providecommand \translation [1]{[#1]}%
\providecommand \BibitemOpen [0]{}%
\providecommand \bibitemStop [0]{}%
\providecommand \bibitemNoStop [0]{.\EOS\space}%
\providecommand \EOS [0]{\spacefactor3000\relax}%
\providecommand \BibitemShut  [1]{\csname bibitem#1\endcsname}%
\let\auto@bib@innerbib\@empty
\bibitem [{\citenamefont {Schaibley}\ \emph {et~al.}(2016)\citenamefont
  {Schaibley}, \citenamefont {Yu}, \citenamefont {Clark}, \citenamefont
  {Rivera}, \citenamefont {Ross}, \citenamefont {Seyler}, \citenamefont {Yao},\
  and\ \citenamefont {Xu}}]{Schaibley2016}%
  \BibitemOpen
  \bibfield  {author} {\bibinfo {author} {\bibfnamefont {J.~R.}\ \bibnamefont
  {Schaibley}}, \bibinfo {author} {\bibfnamefont {H.}~\bibnamefont {Yu}},
  \bibinfo {author} {\bibfnamefont {G.}~\bibnamefont {Clark}}, \bibinfo
  {author} {\bibfnamefont {P.}~\bibnamefont {Rivera}}, \bibinfo {author}
  {\bibfnamefont {J.~S.}\ \bibnamefont {Ross}}, \bibinfo {author}
  {\bibfnamefont {K.~L.}\ \bibnamefont {Seyler}}, \bibinfo {author}
  {\bibfnamefont {W.}~\bibnamefont {Yao}},\ and\ \bibinfo {author}
  {\bibfnamefont {X.}~\bibnamefont {Xu}},\ }\bibfield  {title} {\bibinfo
  {title} {Valleytronics in 2d materials},\ }\bibfield  {journal} {\bibinfo
  {journal} {Nat. Rev. Mat.}\ }\textbf {\bibinfo {volume} {1}},\ \href
  {https://doi.org/10.1038/natrevmats.2016.55} {10.1038/natrevmats.2016.55}
  (\bibinfo {year} {2016})\BibitemShut {NoStop}%
\bibitem [{\citenamefont {Wang}\ \emph {et~al.}(2014)\citenamefont {Wang},
  \citenamefont {Lin},\ and\ \citenamefont {Chan}}]{Wang2014}%
  \BibitemOpen
  \bibfield  {author} {\bibinfo {author} {\bibfnamefont {J.}~\bibnamefont
  {Wang}}, \bibinfo {author} {\bibfnamefont {Z.}~\bibnamefont {Lin}},\ and\
  \bibinfo {author} {\bibfnamefont {K.~S.}\ \bibnamefont {Chan}},\ }\bibfield
  {title} {\bibinfo {title} {Pure valley current generation in graphene with a
  dirac gap by quantum pumping},\ }\href
  {https://doi.org/10.7567/apex.7.125102} {\bibfield  {journal} {\bibinfo
  {journal} {Appl. Phys. Express}\ }\textbf {\bibinfo {volume} {7}},\ \bibinfo
  {pages} {125102} (\bibinfo {year} {2014})}\BibitemShut {NoStop}%
\bibitem [{\citenamefont {Settnes}\ \emph {et~al.}(2016)\citenamefont
  {Settnes}, \citenamefont {Power}, \citenamefont {Brandbyge},\ and\
  \citenamefont {Jauho}}]{Settnes2016}%
  \BibitemOpen
  \bibfield  {author} {\bibinfo {author} {\bibfnamefont {M.}~\bibnamefont
  {Settnes}}, \bibinfo {author} {\bibfnamefont {S.~R.}\ \bibnamefont {Power}},
  \bibinfo {author} {\bibfnamefont {M.}~\bibnamefont {Brandbyge}},\ and\
  \bibinfo {author} {\bibfnamefont {A.-P.}\ \bibnamefont {Jauho}},\ }\bibfield
  {title} {\bibinfo {title} {Graphene nanobubbles as valley filters and beam
  splitters},\ }\bibfield  {journal} {\bibinfo  {journal} {Physical Review
  Letters}\ }\textbf {\bibinfo {volume} {117}},\ \href
  {https://doi.org/10.1103/physrevlett.117.276801}
  {10.1103/physrevlett.117.276801} (\bibinfo {year} {2016})\BibitemShut
  {NoStop}%
\bibitem [{\citenamefont {Stegmann}\ and\ \citenamefont
  {Szpak}(2019)}]{Stegmann2019}%
  \BibitemOpen
  \bibfield  {author} {\bibinfo {author} {\bibfnamefont {T.}~\bibnamefont
  {Stegmann}}\ and\ \bibinfo {author} {\bibfnamefont {N.}~\bibnamefont
  {Szpak}},\ }\bibfield  {title} {\bibinfo {title} {Current splitting and
  valley polarization in elastically deformed graphene},\ }\href
  {https://doi.org/10.1088/2053-1583/aaea8d} {\bibfield  {journal} {\bibinfo
  {journal} {2D Mater.}\ }\textbf {\bibinfo {volume} {6}},\ \bibinfo {pages}
  {015024} (\bibinfo {year} {2019})}\BibitemShut {NoStop}%
\bibitem [{\citenamefont {Milovanovi{\'{c}}}\ and\ \citenamefont
  {Peeters}(2016)}]{Milovanovic2016}%
  \BibitemOpen
  \bibfield  {author} {\bibinfo {author} {\bibfnamefont {S.~P.}\ \bibnamefont
  {Milovanovi{\'{c}}}}\ and\ \bibinfo {author} {\bibfnamefont {F.~M.}\
  \bibnamefont {Peeters}},\ }\bibfield  {title} {\bibinfo {title} {Strain
  controlled valley filtering in multi-terminal graphene structures},\ }\href
  {https://doi.org/10.1063/1.4967977} {\bibfield  {journal} {\bibinfo
  {journal} {Applied Physics Letters}\ }\textbf {\bibinfo {volume} {109}},\
  \bibinfo {pages} {203108} (\bibinfo {year} {2016})}\BibitemShut {NoStop}%
\bibitem [{\citenamefont {Carrillo-Bastos}\ \emph {et~al.}(2016)\citenamefont
  {Carrillo-Bastos}, \citenamefont {Le{\'{o}}n}, \citenamefont {Faria},
  \citenamefont {Latg{\'{e}}}, \citenamefont {Andrei},\ and\ \citenamefont
  {Sandler}}]{Carrillo-Bastos2016}%
  \BibitemOpen
  \bibfield  {author} {\bibinfo {author} {\bibfnamefont {R.}~\bibnamefont
  {Carrillo-Bastos}}, \bibinfo {author} {\bibfnamefont {C.}~\bibnamefont
  {Le{\'{o}}n}}, \bibinfo {author} {\bibfnamefont {D.}~\bibnamefont {Faria}},
  \bibinfo {author} {\bibfnamefont {A.}~\bibnamefont {Latg{\'{e}}}}, \bibinfo
  {author} {\bibfnamefont {E.~Y.}\ \bibnamefont {Andrei}},\ and\ \bibinfo
  {author} {\bibfnamefont {N.}~\bibnamefont {Sandler}},\ }\bibfield  {title}
  {\bibinfo {title} {Strained fold-assisted transport in graphene systems},\
  }\bibfield  {journal} {\bibinfo  {journal} {Physical Review B}\ }\textbf
  {\bibinfo {volume} {94}},\ \href {https://doi.org/10.1103/physrevb.94.125422}
  {10.1103/physrevb.94.125422} (\bibinfo {year} {2016})\BibitemShut {NoStop}%
\bibitem [{\citenamefont {Zhai}\ and\ \citenamefont
  {Sandler}(2018)}]{Zhai2018}%
  \BibitemOpen
  \bibfield  {author} {\bibinfo {author} {\bibfnamefont {D.}~\bibnamefont
  {Zhai}}\ and\ \bibinfo {author} {\bibfnamefont {N.}~\bibnamefont {Sandler}},\
  }\bibfield  {title} {\bibinfo {title} {Local versus extended deformed
  graphene geometries for valley filtering},\ }\bibfield  {journal} {\bibinfo
  {journal} {Physical Review B}\ }\textbf {\bibinfo {volume} {98}},\ \href
  {https://doi.org/10.1103/physrevb.98.165437} {10.1103/physrevb.98.165437}
  (\bibinfo {year} {2018})\BibitemShut {NoStop}%
\bibitem [{\citenamefont {Carrillo-Bastos}\ \emph {et~al.}(2018)\citenamefont
  {Carrillo-Bastos}, \citenamefont {Ochoa}, \citenamefont {Zavala},\ and\
  \citenamefont {Mireles}}]{Carrillo-Bastos2018}%
  \BibitemOpen
  \bibfield  {author} {\bibinfo {author} {\bibfnamefont {R.}~\bibnamefont
  {Carrillo-Bastos}}, \bibinfo {author} {\bibfnamefont {M.}~\bibnamefont
  {Ochoa}}, \bibinfo {author} {\bibfnamefont {S.~A.}\ \bibnamefont {Zavala}},\
  and\ \bibinfo {author} {\bibfnamefont {F.}~\bibnamefont {Mireles}},\
  }\bibfield  {title} {\bibinfo {title} {Enhanced asymmetric valley scattering
  by scalar fields in nonuniform out-of-plane deformations in graphene},\
  }\bibfield  {journal} {\bibinfo  {journal} {Physical Review B}\ }\textbf
  {\bibinfo {volume} {98}},\ \href {https://doi.org/10.1103/physrevb.98.165436}
  {10.1103/physrevb.98.165436} (\bibinfo {year} {2018})\BibitemShut {NoStop}%
\bibitem [{\citenamefont {Mahmud}\ and\ \citenamefont
  {Sandler}(2020)}]{Mahmud2020}%
  \BibitemOpen
  \bibfield  {author} {\bibinfo {author} {\bibfnamefont {M.~T.}\ \bibnamefont
  {Mahmud}}\ and\ \bibinfo {author} {\bibfnamefont {N.}~\bibnamefont
  {Sandler}},\ }\bibfield  {title} {\bibinfo {title} {Emergence of
  strain-induced moir{\'{e}} patterns and pseudomagnetic field confined states
  in graphene},\ }\bibfield  {journal} {\bibinfo  {journal} {Physical Review
  B}\ }\textbf {\bibinfo {volume} {102}},\ \href
  {https://doi.org/10.1103/physrevb.102.235410} {10.1103/physrevb.102.235410}
  (\bibinfo {year} {2020})\BibitemShut {NoStop}%
\bibitem [{\citenamefont {Solomon}\ and\ \citenamefont
  {Power}(2021)}]{Solomon2021}%
  \BibitemOpen
  \bibfield  {author} {\bibinfo {author} {\bibfnamefont {F.}~\bibnamefont
  {Solomon}}\ and\ \bibinfo {author} {\bibfnamefont {S.~R.}\ \bibnamefont
  {Power}},\ }\bibfield  {title} {\bibinfo {title} {Valley current generation
  using biased bilayer graphene dots},\ }\href
  {https://doi.org/10.1103/physrevb.103.235435} {\bibfield  {journal} {\bibinfo
   {journal} {Physical Review B}\ }\textbf {\bibinfo {volume} {103}},\ \bibinfo
  {pages} {235435} (\bibinfo {year} {2021})}\BibitemShut {NoStop}%
\bibitem [{\citenamefont {Lee}\ \emph {et~al.}(2012)\citenamefont {Lee},
  \citenamefont {Lue}, \citenamefont {Wen},\ and\ \citenamefont
  {Wu}}]{Lee2012}%
  \BibitemOpen
  \bibfield  {author} {\bibinfo {author} {\bibfnamefont {M.-K.}\ \bibnamefont
  {Lee}}, \bibinfo {author} {\bibfnamefont {N.-Y.}\ \bibnamefont {Lue}},
  \bibinfo {author} {\bibfnamefont {C.-K.}\ \bibnamefont {Wen}},\ and\ \bibinfo
  {author} {\bibfnamefont {G.~Y.}\ \bibnamefont {Wu}},\ }\bibfield  {title}
  {\bibinfo {title} {Valley-based field-effect transistors in graphene},\
  }\href {https://doi.org/10.1103/physrevb.86.165411} {\bibfield  {journal}
  {\bibinfo  {journal} {Physical Review B}\ }\textbf {\bibinfo {volume} {86}},\
  \bibinfo {pages} {165411} (\bibinfo {year} {2012})}\BibitemShut {NoStop}%
\bibitem [{\citenamefont {Faria}\ \emph {et~al.}(2020)\citenamefont {Faria},
  \citenamefont {Le{\'{o}}n}, \citenamefont {Lima}, \citenamefont
  {Latg{\'{e}}},\ and\ \citenamefont {Sandler}}]{Faria2020}%
  \BibitemOpen
  \bibfield  {author} {\bibinfo {author} {\bibfnamefont {D.}~\bibnamefont
  {Faria}}, \bibinfo {author} {\bibfnamefont {C.}~\bibnamefont {Le{\'{o}}n}},
  \bibinfo {author} {\bibfnamefont {L.~R.~F.}\ \bibnamefont {Lima}}, \bibinfo
  {author} {\bibfnamefont {A.}~\bibnamefont {Latg{\'{e}}}},\ and\ \bibinfo
  {author} {\bibfnamefont {N.}~\bibnamefont {Sandler}},\ }\bibfield  {title}
  {\bibinfo {title} {Valley polarization braiding in strained graphene},\
  }\bibfield  {journal} {\bibinfo  {journal} {Physical Review B}\ }\textbf
  {\bibinfo {volume} {101}},\ \href
  {https://doi.org/10.1103/physrevb.101.081410} {10.1103/physrevb.101.081410}
  (\bibinfo {year} {2020})\BibitemShut {NoStop}%
\bibitem [{\citenamefont {Chen}\ and\ \citenamefont {Hone}(2013)}]{Chen2013b}%
  \BibitemOpen
  \bibfield  {author} {\bibinfo {author} {\bibfnamefont {C.}~\bibnamefont
  {Chen}}\ and\ \bibinfo {author} {\bibfnamefont {J.}~\bibnamefont {Hone}},\
  }\bibfield  {title} {\bibinfo {title} {Graphene nanoelectromechanical
  systems},\ }\href {https://doi.org/10.1109/jproc.2013.2253291} {\bibfield
  {journal} {\bibinfo  {journal} {Proceedings of the {IEEE}}\ }\textbf
  {\bibinfo {volume} {101}},\ \bibinfo {pages} {1766} (\bibinfo {year}
  {2013})}\BibitemShut {NoStop}%
\bibitem [{\citenamefont {Castellanos-Gomez}\ \emph {et~al.}(2014)\citenamefont
  {Castellanos-Gomez}, \citenamefont {Singh}, \citenamefont {van~der Zant},\
  and\ \citenamefont {Steele}}]{Castellanos-Gomez2014}%
  \BibitemOpen
  \bibfield  {author} {\bibinfo {author} {\bibfnamefont {A.}~\bibnamefont
  {Castellanos-Gomez}}, \bibinfo {author} {\bibfnamefont {V.}~\bibnamefont
  {Singh}}, \bibinfo {author} {\bibfnamefont {H.~S.~J.}\ \bibnamefont {van~der
  Zant}},\ and\ \bibinfo {author} {\bibfnamefont {G.~A.}\ \bibnamefont
  {Steele}},\ }\bibfield  {title} {\bibinfo {title} {Mechanics of
  freely-suspended ultrathin layered materials},\ }\href
  {https://doi.org/10.1002/andp.201400153} {\bibfield  {journal} {\bibinfo
  {journal} {Annalen der Physik}\ }\textbf {\bibinfo {volume} {527}},\ \bibinfo
  {pages} {27} (\bibinfo {year} {2014})}\BibitemShut {NoStop}%
\bibitem [{\citenamefont {Khan}\ \emph {et~al.}(2017)\citenamefont {Khan},
  \citenamefont {Kermany}, \citenamefont {Öchsner},\ and\ \citenamefont
  {Iacopi}}]{Khan2017}%
  \BibitemOpen
  \bibfield  {author} {\bibinfo {author} {\bibfnamefont {Z.~H.}\ \bibnamefont
  {Khan}}, \bibinfo {author} {\bibfnamefont {A.~R.}\ \bibnamefont {Kermany}},
  \bibinfo {author} {\bibfnamefont {A.}~\bibnamefont {Öchsner}},\ and\
  \bibinfo {author} {\bibfnamefont {F.}~\bibnamefont {Iacopi}},\ }\bibfield
  {title} {\bibinfo {title} {Mechanical and electromechanical properties of
  graphene and their potential application in {MEMS}},\ }\href
  {https://doi.org/10.1088/1361-6463/50/5/053003} {\bibfield  {journal}
  {\bibinfo  {journal} {Journal of Physics D: Applied Physics}\ }\textbf
  {\bibinfo {volume} {50}},\ \bibinfo {pages} {053003} (\bibinfo {year}
  {2017})}\BibitemShut {NoStop}%
\bibitem [{\citenamefont {Lemme}\ \emph {et~al.}(2020)\citenamefont {Lemme},
  \citenamefont {Wagner}, \citenamefont {Lee}, \citenamefont {Fan},
  \citenamefont {Verbiest}, \citenamefont {Wittmann}, \citenamefont {Lukas},
  \citenamefont {Dolleman}, \citenamefont {Niklaus}, \citenamefont {van~der
  Zant}, \citenamefont {Duesberg},\ and\ \citenamefont
  {Steeneken}}]{Lemme2020}%
  \BibitemOpen
  \bibfield  {author} {\bibinfo {author} {\bibfnamefont {M.~C.}\ \bibnamefont
  {Lemme}}, \bibinfo {author} {\bibfnamefont {S.}~\bibnamefont {Wagner}},
  \bibinfo {author} {\bibfnamefont {K.}~\bibnamefont {Lee}}, \bibinfo {author}
  {\bibfnamefont {X.}~\bibnamefont {Fan}}, \bibinfo {author} {\bibfnamefont
  {G.~J.}\ \bibnamefont {Verbiest}}, \bibinfo {author} {\bibfnamefont
  {S.}~\bibnamefont {Wittmann}}, \bibinfo {author} {\bibfnamefont
  {S.}~\bibnamefont {Lukas}}, \bibinfo {author} {\bibfnamefont {R.~J.}\
  \bibnamefont {Dolleman}}, \bibinfo {author} {\bibfnamefont {F.}~\bibnamefont
  {Niklaus}}, \bibinfo {author} {\bibfnamefont {H.~S.~J.}\ \bibnamefont
  {van~der Zant}}, \bibinfo {author} {\bibfnamefont {G.~S.}\ \bibnamefont
  {Duesberg}},\ and\ \bibinfo {author} {\bibfnamefont {P.~G.}\ \bibnamefont
  {Steeneken}},\ }\bibfield  {title} {\bibinfo {title} {Nanoelectromechanical
  sensors based on suspended 2d materials},\ }\href
  {https://doi.org/10.34133/2020/8748602} {\bibfield  {journal} {\bibinfo
  {journal} {Research}\ }\textbf {\bibinfo {volume} {2020}},\ \bibinfo {pages}
  {1} (\bibinfo {year} {2020})}\BibitemShut {NoStop}%
\bibitem [{\citenamefont {Chen}\ \emph {et~al.}(2009)\citenamefont {Chen},
  \citenamefont {Rosenblatt}, \citenamefont {Bolotin}, \citenamefont {Kalb},
  \citenamefont {Kim}, \citenamefont {Kymissis}, \citenamefont {Stormer},
  \citenamefont {Heinz},\ and\ \citenamefont {Hone}}]{Chen2009}%
  \BibitemOpen
  \bibfield  {author} {\bibinfo {author} {\bibfnamefont {C.}~\bibnamefont
  {Chen}}, \bibinfo {author} {\bibfnamefont {S.}~\bibnamefont {Rosenblatt}},
  \bibinfo {author} {\bibfnamefont {K.~I.}\ \bibnamefont {Bolotin}}, \bibinfo
  {author} {\bibfnamefont {W.}~\bibnamefont {Kalb}}, \bibinfo {author}
  {\bibfnamefont {P.}~\bibnamefont {Kim}}, \bibinfo {author} {\bibfnamefont
  {I.}~\bibnamefont {Kymissis}}, \bibinfo {author} {\bibfnamefont {H.~L.}\
  \bibnamefont {Stormer}}, \bibinfo {author} {\bibfnamefont {T.~F.}\
  \bibnamefont {Heinz}},\ and\ \bibinfo {author} {\bibfnamefont
  {J.}~\bibnamefont {Hone}},\ }\bibfield  {title} {\bibinfo {title}
  {Performance of monolayer graphene nanomechanical resonators with electrical
  readout},\ }\href {https://doi.org/10.1038/nnano.2009.267} {\bibfield
  {journal} {\bibinfo  {journal} {Nature Nanotechnology}\ }\textbf {\bibinfo
  {volume} {4}},\ \bibinfo {pages} {861} (\bibinfo {year} {2009})}\BibitemShut
  {NoStop}%
\bibitem [{\citenamefont {Wong}\ \emph {et~al.}(2010)\citenamefont {Wong},
  \citenamefont {Annamalai}, \citenamefont {Wang},\ and\ \citenamefont
  {Palaniapan}}]{Wong2010}%
  \BibitemOpen
  \bibfield  {author} {\bibinfo {author} {\bibfnamefont {C.-L.}\ \bibnamefont
  {Wong}}, \bibinfo {author} {\bibfnamefont {M.}~\bibnamefont {Annamalai}},
  \bibinfo {author} {\bibfnamefont {Z.-Q.}\ \bibnamefont {Wang}},\ and\
  \bibinfo {author} {\bibfnamefont {M.}~\bibnamefont {Palaniapan}},\ }\bibfield
   {title} {\bibinfo {title} {Characterization of nanomechanical graphene drum
  structures},\ }\href {https://doi.org/10.1088/0960-1317/20/11/115029}
  {\bibfield  {journal} {\bibinfo  {journal} {Journal of Micromechanics and
  Microengineering}\ }\textbf {\bibinfo {volume} {20}},\ \bibinfo {pages}
  {115029} (\bibinfo {year} {2010})}\BibitemShut {NoStop}%
\bibitem [{\citenamefont {Xu}\ \emph {et~al.}(2010)\citenamefont {Xu},
  \citenamefont {Chen}, \citenamefont {Deshpande}, \citenamefont {DiRenno},
  \citenamefont {Gondarenko}, \citenamefont {Heinz}, \citenamefont {Liu},
  \citenamefont {Kim},\ and\ \citenamefont {Hone}}]{Xu2010}%
  \BibitemOpen
  \bibfield  {author} {\bibinfo {author} {\bibfnamefont {Y.}~\bibnamefont
  {Xu}}, \bibinfo {author} {\bibfnamefont {C.}~\bibnamefont {Chen}}, \bibinfo
  {author} {\bibfnamefont {V.~V.}\ \bibnamefont {Deshpande}}, \bibinfo {author}
  {\bibfnamefont {F.~A.}\ \bibnamefont {DiRenno}}, \bibinfo {author}
  {\bibfnamefont {A.}~\bibnamefont {Gondarenko}}, \bibinfo {author}
  {\bibfnamefont {D.~B.}\ \bibnamefont {Heinz}}, \bibinfo {author}
  {\bibfnamefont {S.}~\bibnamefont {Liu}}, \bibinfo {author} {\bibfnamefont
  {P.}~\bibnamefont {Kim}},\ and\ \bibinfo {author} {\bibfnamefont
  {J.}~\bibnamefont {Hone}},\ }\bibfield  {title} {\bibinfo {title} {Radio
  frequency electrical transduction of graphene mechanical resonators},\ }\href
  {https://doi.org/10.1063/1.3528341} {\bibfield  {journal} {\bibinfo
  {journal} {Applied Physics Letters}\ }\textbf {\bibinfo {volume} {97}},\
  \bibinfo {pages} {243111} (\bibinfo {year} {2010})}\BibitemShut {NoStop}%
\bibitem [{\citenamefont {Lee}\ \emph {et~al.}(2013)\citenamefont {Lee},
  \citenamefont {Chen}, \citenamefont {Deshpande}, \citenamefont {Lee},
  \citenamefont {Lee}, \citenamefont {Lekas}, \citenamefont {Gondarenko},
  \citenamefont {Yu}, \citenamefont {Shepard}, \citenamefont {Kim},\ and\
  \citenamefont {Hone}}]{Lee2013}%
  \BibitemOpen
  \bibfield  {author} {\bibinfo {author} {\bibfnamefont {S.}~\bibnamefont
  {Lee}}, \bibinfo {author} {\bibfnamefont {C.}~\bibnamefont {Chen}}, \bibinfo
  {author} {\bibfnamefont {V.~V.}\ \bibnamefont {Deshpande}}, \bibinfo {author}
  {\bibfnamefont {G.-H.}\ \bibnamefont {Lee}}, \bibinfo {author} {\bibfnamefont
  {I.}~\bibnamefont {Lee}}, \bibinfo {author} {\bibfnamefont {M.}~\bibnamefont
  {Lekas}}, \bibinfo {author} {\bibfnamefont {A.}~\bibnamefont {Gondarenko}},
  \bibinfo {author} {\bibfnamefont {Y.-J.}\ \bibnamefont {Yu}}, \bibinfo
  {author} {\bibfnamefont {K.}~\bibnamefont {Shepard}}, \bibinfo {author}
  {\bibfnamefont {P.}~\bibnamefont {Kim}},\ and\ \bibinfo {author}
  {\bibfnamefont {J.}~\bibnamefont {Hone}},\ }\bibfield  {title} {\bibinfo
  {title} {Electrically integrated {SU}-8 clamped graphene drum resonators for
  strain engineering},\ }\href {https://doi.org/10.1063/1.4793302} {\bibfield
  {journal} {\bibinfo  {journal} {Applied Physics Letters}\ }\textbf {\bibinfo
  {volume} {102}},\ \bibinfo {pages} {153101} (\bibinfo {year}
  {2013})}\BibitemShut {NoStop}%
\bibitem [{\citenamefont {Mathew}\ \emph {et~al.}(2016)\citenamefont {Mathew},
  \citenamefont {Patel}, \citenamefont {Borah}, \citenamefont {Vijay},\ and\
  \citenamefont {Deshmukh}}]{Mathew2016}%
  \BibitemOpen
  \bibfield  {author} {\bibinfo {author} {\bibfnamefont {J.~P.}\ \bibnamefont
  {Mathew}}, \bibinfo {author} {\bibfnamefont {R.~N.}\ \bibnamefont {Patel}},
  \bibinfo {author} {\bibfnamefont {A.}~\bibnamefont {Borah}}, \bibinfo
  {author} {\bibfnamefont {R.}~\bibnamefont {Vijay}},\ and\ \bibinfo {author}
  {\bibfnamefont {M.~M.}\ \bibnamefont {Deshmukh}},\ }\bibfield  {title}
  {\bibinfo {title} {Dynamical strong coupling and parametric amplification of
  mechanical modes of graphene drums},\ }\href
  {https://doi.org/10.1038/nnano.2016.94} {\bibfield  {journal} {\bibinfo
  {journal} {Nature Nanotechnology}\ }\textbf {\bibinfo {volume} {11}},\
  \bibinfo {pages} {747} (\bibinfo {year} {2016})}\BibitemShut {NoStop}%
\bibitem [{\citenamefont {Alba}\ \emph {et~al.}(2016)\citenamefont {Alba},
  \citenamefont {Massel}, \citenamefont {Storch}, \citenamefont {Abhilash},
  \citenamefont {Hui}, \citenamefont {McEuen}, \citenamefont {Craighead},\ and\
  \citenamefont {Parpia}}]{Alba2016}%
  \BibitemOpen
  \bibfield  {author} {\bibinfo {author} {\bibfnamefont {R.~D.}\ \bibnamefont
  {Alba}}, \bibinfo {author} {\bibfnamefont {F.}~\bibnamefont {Massel}},
  \bibinfo {author} {\bibfnamefont {I.~R.}\ \bibnamefont {Storch}}, \bibinfo
  {author} {\bibfnamefont {T.~S.}\ \bibnamefont {Abhilash}}, \bibinfo {author}
  {\bibfnamefont {A.}~\bibnamefont {Hui}}, \bibinfo {author} {\bibfnamefont
  {P.~L.}\ \bibnamefont {McEuen}}, \bibinfo {author} {\bibfnamefont {H.~G.}\
  \bibnamefont {Craighead}},\ and\ \bibinfo {author} {\bibfnamefont {J.~M.}\
  \bibnamefont {Parpia}},\ }\bibfield  {title} {\bibinfo {title} {Tunable
  phonon-cavity coupling in graphene membranes},\ }\href
  {https://doi.org/10.1038/nnano.2016.86} {\bibfield  {journal} {\bibinfo
  {journal} {Nature Nanotechnology}\ }\textbf {\bibinfo {volume} {11}},\
  \bibinfo {pages} {741} (\bibinfo {year} {2016})}\BibitemShut {NoStop}%
\bibitem [{\citenamefont {Davidovikj}\ \emph {et~al.}(2016)\citenamefont
  {Davidovikj}, \citenamefont {Slim}, \citenamefont {Cartamil-Bueno},
  \citenamefont {van~der Zant}, \citenamefont {Steeneken},\ and\ \citenamefont
  {Venstra}}]{Davidovikj2016}%
  \BibitemOpen
  \bibfield  {author} {\bibinfo {author} {\bibfnamefont {D.}~\bibnamefont
  {Davidovikj}}, \bibinfo {author} {\bibfnamefont {J.~J.}\ \bibnamefont
  {Slim}}, \bibinfo {author} {\bibfnamefont {S.~J.}\ \bibnamefont
  {Cartamil-Bueno}}, \bibinfo {author} {\bibfnamefont {H.~S.~J.}\ \bibnamefont
  {van~der Zant}}, \bibinfo {author} {\bibfnamefont {P.~G.}\ \bibnamefont
  {Steeneken}},\ and\ \bibinfo {author} {\bibfnamefont {W.~J.}\ \bibnamefont
  {Venstra}},\ }\bibfield  {title} {\bibinfo {title} {Visualizing the motion of
  graphene nanodrums},\ }\href {https://doi.org/10.1021/acs.nanolett.6b00477}
  {\bibfield  {journal} {\bibinfo  {journal} {Nano Letters}\ }\textbf {\bibinfo
  {volume} {16}},\ \bibinfo {pages} {2768} (\bibinfo {year}
  {2016})}\BibitemShut {NoStop}%
\bibitem [{\citenamefont {Davidovikj}\ \emph {et~al.}(2018)\citenamefont
  {Davidovikj}, \citenamefont {Poot}, \citenamefont {Cartamil-Bueno},
  \citenamefont {van~der Zant},\ and\ \citenamefont
  {Steeneken}}]{Davidovikj2018}%
  \BibitemOpen
  \bibfield  {author} {\bibinfo {author} {\bibfnamefont {D.}~\bibnamefont
  {Davidovikj}}, \bibinfo {author} {\bibfnamefont {M.}~\bibnamefont {Poot}},
  \bibinfo {author} {\bibfnamefont {S.~J.}\ \bibnamefont {Cartamil-Bueno}},
  \bibinfo {author} {\bibfnamefont {H.~S.~J.}\ \bibnamefont {van~der Zant}},\
  and\ \bibinfo {author} {\bibfnamefont {P.~G.}\ \bibnamefont {Steeneken}},\
  }\bibfield  {title} {\bibinfo {title} {On-chip heaters for tension tuning of
  graphene nanodrums},\ }\href {https://doi.org/10.1021/acs.nanolett.7b05358}
  {\bibfield  {journal} {\bibinfo  {journal} {Nano Letters}\ }\textbf {\bibinfo
  {volume} {18}},\ \bibinfo {pages} {2852} (\bibinfo {year}
  {2018})}\BibitemShut {NoStop}%
\bibitem [{\citenamefont {Güttinger}\ \emph {et~al.}(2017)\citenamefont
  {Güttinger}, \citenamefont {Noury}, \citenamefont {Weber}, \citenamefont
  {Eriksson}, \citenamefont {Lagoin}, \citenamefont {Moser}, \citenamefont
  {Eichler}, \citenamefont {Wallraff}, \citenamefont {Isacsson},\ and\
  \citenamefont {Bachtold}}]{Guettinger2017}%
  \BibitemOpen
  \bibfield  {author} {\bibinfo {author} {\bibfnamefont {J.}~\bibnamefont
  {Güttinger}}, \bibinfo {author} {\bibfnamefont {A.}~\bibnamefont {Noury}},
  \bibinfo {author} {\bibfnamefont {P.}~\bibnamefont {Weber}}, \bibinfo
  {author} {\bibfnamefont {A.~M.}\ \bibnamefont {Eriksson}}, \bibinfo {author}
  {\bibfnamefont {C.}~\bibnamefont {Lagoin}}, \bibinfo {author} {\bibfnamefont
  {J.}~\bibnamefont {Moser}}, \bibinfo {author} {\bibfnamefont
  {C.}~\bibnamefont {Eichler}}, \bibinfo {author} {\bibfnamefont
  {A.}~\bibnamefont {Wallraff}}, \bibinfo {author} {\bibfnamefont
  {A.}~\bibnamefont {Isacsson}},\ and\ \bibinfo {author} {\bibfnamefont
  {A.}~\bibnamefont {Bachtold}},\ }\bibfield  {title} {\bibinfo {title}
  {Energy-dependent path of dissipation in nanomechanical resonators},\ }\href
  {https://doi.org/10.1038/nnano.2017.86} {\bibfield  {journal} {\bibinfo
  {journal} {Nature Nanotechnology}\ }\textbf {\bibinfo {volume} {12}},\
  \bibinfo {pages} {631} (\bibinfo {year} {2017})}\BibitemShut {NoStop}%
\bibitem [{\citenamefont {Shin}\ \emph {et~al.}(2016)\citenamefont {Shin},
  \citenamefont {Lozada-Hidalgo}, \citenamefont {Sambricio}, \citenamefont
  {Grigorieva}, \citenamefont {Geim},\ and\ \citenamefont
  {Casiraghi}}]{Shin2016}%
  \BibitemOpen
  \bibfield  {author} {\bibinfo {author} {\bibfnamefont {Y.}~\bibnamefont
  {Shin}}, \bibinfo {author} {\bibfnamefont {M.}~\bibnamefont
  {Lozada-Hidalgo}}, \bibinfo {author} {\bibfnamefont {J.~L.}\ \bibnamefont
  {Sambricio}}, \bibinfo {author} {\bibfnamefont {I.~V.}\ \bibnamefont
  {Grigorieva}}, \bibinfo {author} {\bibfnamefont {A.~K.}\ \bibnamefont
  {Geim}},\ and\ \bibinfo {author} {\bibfnamefont {C.}~\bibnamefont
  {Casiraghi}},\ }\bibfield  {title} {\bibinfo {title} {Raman spectroscopy of
  highly pressurized graphene membranes},\ }\href
  {https://doi.org/10.1063/1.4952972} {\bibfield  {journal} {\bibinfo
  {journal} {Applied Physics Letters}\ }\textbf {\bibinfo {volume} {108}},\
  \bibinfo {pages} {221907} (\bibinfo {year} {2016})}\BibitemShut {NoStop}%
\bibitem [{\citenamefont {Smith}\ \emph {et~al.}(2016)\citenamefont {Smith},
  \citenamefont {Niklaus}, \citenamefont {Paussa}, \citenamefont {Schröder},
  \citenamefont {Fischer}, \citenamefont {Sterner}, \citenamefont {Wagner},
  \citenamefont {Vaziri}, \citenamefont {Forsberg}, \citenamefont {Esseni},
  \citenamefont {Östling},\ and\ \citenamefont {Lemme}}]{Smith2016}%
  \BibitemOpen
  \bibfield  {author} {\bibinfo {author} {\bibfnamefont {A.~D.}\ \bibnamefont
  {Smith}}, \bibinfo {author} {\bibfnamefont {F.}~\bibnamefont {Niklaus}},
  \bibinfo {author} {\bibfnamefont {A.}~\bibnamefont {Paussa}}, \bibinfo
  {author} {\bibfnamefont {S.}~\bibnamefont {Schröder}}, \bibinfo {author}
  {\bibfnamefont {A.~C.}\ \bibnamefont {Fischer}}, \bibinfo {author}
  {\bibfnamefont {M.}~\bibnamefont {Sterner}}, \bibinfo {author} {\bibfnamefont
  {S.}~\bibnamefont {Wagner}}, \bibinfo {author} {\bibfnamefont
  {S.}~\bibnamefont {Vaziri}}, \bibinfo {author} {\bibfnamefont
  {F.}~\bibnamefont {Forsberg}}, \bibinfo {author} {\bibfnamefont
  {D.}~\bibnamefont {Esseni}}, \bibinfo {author} {\bibfnamefont
  {M.}~\bibnamefont {Östling}},\ and\ \bibinfo {author} {\bibfnamefont
  {M.~C.}\ \bibnamefont {Lemme}},\ }\bibfield  {title} {\bibinfo {title}
  {Piezoresistive properties of suspended graphene membranes under uniaxial and
  biaxial strain in nanoelectromechanical pressure sensors},\ }\href
  {https://doi.org/10.1021/acsnano.6b02533} {\bibfield  {journal} {\bibinfo
  {journal} {{ACS} Nano}\ }\textbf {\bibinfo {volume} {10}},\ \bibinfo {pages}
  {9879} (\bibinfo {year} {2016})}\BibitemShut {NoStop}%
\bibitem [{\citenamefont {Atalaya}\ \emph {et~al.}(2008)\citenamefont
  {Atalaya}, \citenamefont {Isacsson},\ and\ \citenamefont
  {Kinaret}}]{Atalaya2008}%
  \BibitemOpen
  \bibfield  {author} {\bibinfo {author} {\bibfnamefont {J.}~\bibnamefont
  {Atalaya}}, \bibinfo {author} {\bibfnamefont {A.}~\bibnamefont {Isacsson}},\
  and\ \bibinfo {author} {\bibfnamefont {J.~M.}\ \bibnamefont {Kinaret}},\
  }\bibfield  {title} {\bibinfo {title} {Continuum elastic modeling of graphene
  resonators},\ }\href {https://doi.org/10.1021/nl801733d} {\bibfield
  {journal} {\bibinfo  {journal} {Nano letters}\ }\textbf {\bibinfo {volume}
  {8}},\ \bibinfo {pages} {4196} (\bibinfo {year} {2008})}\BibitemShut
  {NoStop}%
\bibitem [{\citenamefont {Eriksson}\ \emph {et~al.}(2013)\citenamefont
  {Eriksson}, \citenamefont {Midtvedt}, \citenamefont {Croy},\ and\
  \citenamefont {Isacsson}}]{Eriksson2013}%
  \BibitemOpen
  \bibfield  {author} {\bibinfo {author} {\bibfnamefont {A.~M.}\ \bibnamefont
  {Eriksson}}, \bibinfo {author} {\bibfnamefont {D.}~\bibnamefont {Midtvedt}},
  \bibinfo {author} {\bibfnamefont {A.}~\bibnamefont {Croy}},\ and\ \bibinfo
  {author} {\bibfnamefont {A.}~\bibnamefont {Isacsson}},\ }\bibfield  {title}
  {\bibinfo {title} {Frequency tuning, nonlinearities and mode coupling in
  circular mechanical graphene resonators},\ }\href
  {https://doi.org/10.1088/0957-4484/24/39/395702} {\bibfield  {journal}
  {\bibinfo  {journal} {Nanotechnology}\ }\textbf {\bibinfo {volume} {24}},\
  \bibinfo {pages} {395702} (\bibinfo {year} {2013})}\BibitemShut {NoStop}%
\bibitem [{\citenamefont {Zhang}\ \emph {et~al.}(2015)\citenamefont {Zhang},
  \citenamefont {Waitz}, \citenamefont {Yang}, \citenamefont {Lutz},
  \citenamefont {Angelova}, \citenamefont {G{\"o}lzh{\"a}user},\ and\
  \citenamefont {Scheer}}]{Zhang2015}%
  \BibitemOpen
  \bibfield  {author} {\bibinfo {author} {\bibfnamefont {X.}~\bibnamefont
  {Zhang}}, \bibinfo {author} {\bibfnamefont {R.}~\bibnamefont {Waitz}},
  \bibinfo {author} {\bibfnamefont {F.}~\bibnamefont {Yang}}, \bibinfo {author}
  {\bibfnamefont {C.}~\bibnamefont {Lutz}}, \bibinfo {author} {\bibfnamefont
  {P.}~\bibnamefont {Angelova}}, \bibinfo {author} {\bibfnamefont
  {A.}~\bibnamefont {G{\"o}lzh{\"a}user}},\ and\ \bibinfo {author}
  {\bibfnamefont {E.}~\bibnamefont {Scheer}},\ }\bibfield  {title} {\bibinfo
  {title} {Vibrational modes of ultrathin carbon nanomembrane mechanical
  resonators},\ }\href {https://doi.org/10.1063/1.4908058} {\bibfield
  {journal} {\bibinfo  {journal} {Applied Physics Letters}\ }\textbf {\bibinfo
  {volume} {106}},\ \bibinfo {pages} {063107} (\bibinfo {year}
  {2015})}\BibitemShut {NoStop}%
\bibitem [{\citenamefont {Popescu}\ and\ \citenamefont
  {Croy}(2016)}]{Popescu2016}%
  \BibitemOpen
  \bibfield  {author} {\bibinfo {author} {\bibfnamefont {B.~S.}\ \bibnamefont
  {Popescu}}\ and\ \bibinfo {author} {\bibfnamefont {A.}~\bibnamefont {Croy}},\
  }\bibfield  {title} {\bibinfo {title} {Efficient auxiliary-mode approach for
  time-dependent nanoelectronics},\ }\href
  {https://doi.org/10.1088/1367-2630/18/9/093044} {\bibfield  {journal}
  {\bibinfo  {journal} {New Journal of Physics}\ }\textbf {\bibinfo {volume}
  {18}},\ \bibinfo {pages} {093044} (\bibinfo {year} {2016})}\BibitemShut
  {NoStop}%
\bibitem [{\citenamefont {Miller}\ and\ \citenamefont
  {Alem{\'{a}}n}(2019)}]{Miller2019}%
  \BibitemOpen
  \bibfield  {author} {\bibinfo {author} {\bibfnamefont {D.}~\bibnamefont
  {Miller}}\ and\ \bibinfo {author} {\bibfnamefont {B.}~\bibnamefont
  {Alem{\'{a}}n}},\ }\bibfield  {title} {\bibinfo {title} {Spatially resolved
  optical excitation of mechanical modes in graphene {NEMS}},\ }\href
  {https://doi.org/10.1063/1.5111755} {\bibfield  {journal} {\bibinfo
  {journal} {Applied Physics Letters}\ }\textbf {\bibinfo {volume} {115}},\
  \bibinfo {pages} {193102} (\bibinfo {year} {2019})}\BibitemShut {NoStop}%
\bibitem [{\citenamefont {Chen}\ \emph {et~al.}(2013)\citenamefont {Chen},
  \citenamefont {Lee}, \citenamefont {Deshpande}, \citenamefont {Lee},
  \citenamefont {Lekas}, \citenamefont {Shepard},\ and\ \citenamefont
  {Hone}}]{Chen2013}%
  \BibitemOpen
  \bibfield  {author} {\bibinfo {author} {\bibfnamefont {C.}~\bibnamefont
  {Chen}}, \bibinfo {author} {\bibfnamefont {S.}~\bibnamefont {Lee}}, \bibinfo
  {author} {\bibfnamefont {V.~V.}\ \bibnamefont {Deshpande}}, \bibinfo {author}
  {\bibfnamefont {G.-H.}\ \bibnamefont {Lee}}, \bibinfo {author} {\bibfnamefont
  {M.}~\bibnamefont {Lekas}}, \bibinfo {author} {\bibfnamefont
  {K.}~\bibnamefont {Shepard}},\ and\ \bibinfo {author} {\bibfnamefont
  {J.}~\bibnamefont {Hone}},\ }\bibfield  {title} {\bibinfo {title} {Graphene
  mechanical oscillators with tunable frequency},\ }\href
  {https://doi.org/10.1038/nnano.2013.232} {\bibfield  {journal} {\bibinfo
  {journal} {Nature Nanotechnology}\ }\textbf {\bibinfo {volume} {8}},\
  \bibinfo {pages} {923} (\bibinfo {year} {2013})}\BibitemShut {NoStop}%
\bibitem [{\citenamefont {Lee}\ \emph {et~al.}(2019)\citenamefont {Lee},
  \citenamefont {Davidovikj}, \citenamefont {Sajadi}, \citenamefont
  {{\v{S}}i{\v{s}}kins}, \citenamefont {Alijani}, \citenamefont {van~der
  Zant},\ and\ \citenamefont {Steeneken}}]{Lee2019}%
  \BibitemOpen
  \bibfield  {author} {\bibinfo {author} {\bibfnamefont {M.}~\bibnamefont
  {Lee}}, \bibinfo {author} {\bibfnamefont {D.}~\bibnamefont {Davidovikj}},
  \bibinfo {author} {\bibfnamefont {B.}~\bibnamefont {Sajadi}}, \bibinfo
  {author} {\bibfnamefont {M.}~\bibnamefont {{\v{S}}i{\v{s}}kins}}, \bibinfo
  {author} {\bibfnamefont {F.}~\bibnamefont {Alijani}}, \bibinfo {author}
  {\bibfnamefont {H.~S.~J.}\ \bibnamefont {van~der Zant}},\ and\ \bibinfo
  {author} {\bibfnamefont {P.~G.}\ \bibnamefont {Steeneken}},\ }\bibfield
  {title} {\bibinfo {title} {Sealing graphene nanodrums},\ }\href
  {https://doi.org/10.1021/acs.nanolett.9b01770} {\bibfield  {journal}
  {\bibinfo  {journal} {Nano Letters}\ }\textbf {\bibinfo {volume} {19}},\
  \bibinfo {pages} {5313} (\bibinfo {year} {2019})}\BibitemShut {NoStop}%
\bibitem [{\citenamefont {Vozmediano}\ \emph {et~al.}(2010)\citenamefont
  {Vozmediano}, \citenamefont {Katsnelson},\ and\ \citenamefont
  {Guinea}}]{Vozmediano2010}%
  \BibitemOpen
  \bibfield  {author} {\bibinfo {author} {\bibfnamefont {M.}~\bibnamefont
  {Vozmediano}}, \bibinfo {author} {\bibfnamefont {M.}~\bibnamefont
  {Katsnelson}},\ and\ \bibinfo {author} {\bibfnamefont {F.}~\bibnamefont
  {Guinea}},\ }\bibfield  {title} {\bibinfo {title} {Gauge fields in
  graphene},\ }\href {https://doi.org/10.1016/j.physrep.2010.07.003} {\bibfield
   {journal} {\bibinfo  {journal} {Physics Reports}\ }\textbf {\bibinfo
  {volume} {496}},\ \bibinfo {pages} {109} (\bibinfo {year}
  {2010})}\BibitemShut {NoStop}%
\bibitem [{\citenamefont {Stegmann}\ and\ \citenamefont
  {Szpak}(2016)}]{Stegmann2016}%
  \BibitemOpen
  \bibfield  {author} {\bibinfo {author} {\bibfnamefont {T.}~\bibnamefont
  {Stegmann}}\ and\ \bibinfo {author} {\bibfnamefont {N.}~\bibnamefont
  {Szpak}},\ }\bibfield  {title} {\bibinfo {title} {Current flow paths in
  deformed graphene: from quantum transport to classical trajectories in curved
  space},\ }\href {https://doi.org/10.1088/1367-2630/18/5/053016} {\bibfield
  {journal} {\bibinfo  {journal} {New J. Phys.}\ }\textbf {\bibinfo {volume}
  {18}},\ \bibinfo {pages} {053016} (\bibinfo {year} {2016})}\BibitemShut
  {NoStop}%
\bibitem [{\citenamefont {Naumis}\ \emph {et~al.}(2017)\citenamefont {Naumis},
  \citenamefont {Barraza-Lopez}, \citenamefont {Oliva-Leyva},\ and\
  \citenamefont {Terrones}}]{Naumis2017}%
  \BibitemOpen
  \bibfield  {author} {\bibinfo {author} {\bibfnamefont {G.~G.}\ \bibnamefont
  {Naumis}}, \bibinfo {author} {\bibfnamefont {S.}~\bibnamefont
  {Barraza-Lopez}}, \bibinfo {author} {\bibfnamefont {M.}~\bibnamefont
  {Oliva-Leyva}},\ and\ \bibinfo {author} {\bibfnamefont {H.}~\bibnamefont
  {Terrones}},\ }\bibfield  {title} {\bibinfo {title} {Electronic and optical
  properties of strained graphene and other strained 2d materials: a review},\
  }\href {https://doi.org/10.1088/1361-6633/aa74ef} {\bibfield  {journal}
  {\bibinfo  {journal} {Reports on Progress in Physics}\ }\textbf {\bibinfo
  {volume} {80}},\ \bibinfo {pages} {096501} (\bibinfo {year}
  {2017})}\BibitemShut {NoStop}%
\bibitem [{\citenamefont {Dai}\ \emph {et~al.}(2019)\citenamefont {Dai},
  \citenamefont {Liu},\ and\ \citenamefont {Zhang}}]{Dai2019}%
  \BibitemOpen
  \bibfield  {author} {\bibinfo {author} {\bibfnamefont {Z.}~\bibnamefont
  {Dai}}, \bibinfo {author} {\bibfnamefont {L.}~\bibnamefont {Liu}},\ and\
  \bibinfo {author} {\bibfnamefont {Z.}~\bibnamefont {Zhang}},\ }\bibfield
  {title} {\bibinfo {title} {Strain engineering of 2d materials: Issues and
  opportunities at the interface},\ }\href
  {https://doi.org/10.1002/adma.201805417} {\bibfield  {journal} {\bibinfo
  {journal} {Advanced Materials}\ }\textbf {\bibinfo {volume} {31}},\ \bibinfo
  {pages} {1805417} (\bibinfo {year} {2019})}\BibitemShut {NoStop}%
\bibitem [{\citenamefont {van~der Zande}\ \emph {et~al.}(2010)\citenamefont
  {van~der Zande}, \citenamefont {Barton}, \citenamefont {Alden}, \citenamefont
  {Ruiz-Vargas}, \citenamefont {Whitney}, \citenamefont {Pham}, \citenamefont
  {Park}, \citenamefont {Parpia}, \citenamefont {Craighead},\ and\
  \citenamefont {McEuen}}]{Zande2010}%
  \BibitemOpen
  \bibfield  {author} {\bibinfo {author} {\bibfnamefont {A.~M.}\ \bibnamefont
  {van~der Zande}}, \bibinfo {author} {\bibfnamefont {R.~A.}\ \bibnamefont
  {Barton}}, \bibinfo {author} {\bibfnamefont {J.~S.}\ \bibnamefont {Alden}},
  \bibinfo {author} {\bibfnamefont {C.~S.}\ \bibnamefont {Ruiz-Vargas}},
  \bibinfo {author} {\bibfnamefont {W.~S.}\ \bibnamefont {Whitney}}, \bibinfo
  {author} {\bibfnamefont {P.~H.~Q.}\ \bibnamefont {Pham}}, \bibinfo {author}
  {\bibfnamefont {J.}~\bibnamefont {Park}}, \bibinfo {author} {\bibfnamefont
  {J.~M.}\ \bibnamefont {Parpia}}, \bibinfo {author} {\bibfnamefont {H.~G.}\
  \bibnamefont {Craighead}},\ and\ \bibinfo {author} {\bibfnamefont {P.~L.}\
  \bibnamefont {McEuen}},\ }\bibfield  {title} {\bibinfo {title} {Large-scale
  arrays of single-layer graphene resonators},\ }\href
  {https://doi.org/10.1021/nl102713c} {\bibfield  {journal} {\bibinfo
  {journal} {Nano Letters}\ }\textbf {\bibinfo {volume} {10}},\ \bibinfo
  {pages} {4869} (\bibinfo {year} {2010})}\BibitemShut {NoStop}%
\bibitem [{\citenamefont {Barton}\ \emph {et~al.}(2011)\citenamefont {Barton},
  \citenamefont {Ilic}, \citenamefont {van~der Zande}, \citenamefont {Whitney},
  \citenamefont {McEuen}, \citenamefont {Parpia},\ and\ \citenamefont
  {Craighead}}]{Barton2011}%
  \BibitemOpen
  \bibfield  {author} {\bibinfo {author} {\bibfnamefont {R.~A.}\ \bibnamefont
  {Barton}}, \bibinfo {author} {\bibfnamefont {B.}~\bibnamefont {Ilic}},
  \bibinfo {author} {\bibfnamefont {A.~M.}\ \bibnamefont {van~der Zande}},
  \bibinfo {author} {\bibfnamefont {W.~S.}\ \bibnamefont {Whitney}}, \bibinfo
  {author} {\bibfnamefont {P.~L.}\ \bibnamefont {McEuen}}, \bibinfo {author}
  {\bibfnamefont {J.~M.}\ \bibnamefont {Parpia}},\ and\ \bibinfo {author}
  {\bibfnamefont {H.~G.}\ \bibnamefont {Craighead}},\ }\bibfield  {title}
  {\bibinfo {title} {High, size-dependent quality factor in an array of
  graphene mechanical resonators},\ }\href {https://doi.org/10.1021/nl1042227}
  {\bibfield  {journal} {\bibinfo  {journal} {Nano Letters}\ }\textbf {\bibinfo
  {volume} {11}},\ \bibinfo {pages} {1232} (\bibinfo {year}
  {2011})}\BibitemShut {NoStop}%
\bibitem [{\citenamefont {Liu}\ \emph {et~al.}(2013)\citenamefont {Liu},
  \citenamefont {Suk}, \citenamefont {Boddeti}, \citenamefont {Cantley},
  \citenamefont {Wang}, \citenamefont {Gray}, \citenamefont {Hall},
  \citenamefont {Bright}, \citenamefont {Rogers}, \citenamefont {Dunn},
  \citenamefont {Ruoff},\ and\ \citenamefont {Bunch}}]{Liu2013}%
  \BibitemOpen
  \bibfield  {author} {\bibinfo {author} {\bibfnamefont {X.}~\bibnamefont
  {Liu}}, \bibinfo {author} {\bibfnamefont {J.~W.}\ \bibnamefont {Suk}},
  \bibinfo {author} {\bibfnamefont {N.~G.}\ \bibnamefont {Boddeti}}, \bibinfo
  {author} {\bibfnamefont {L.}~\bibnamefont {Cantley}}, \bibinfo {author}
  {\bibfnamefont {L.}~\bibnamefont {Wang}}, \bibinfo {author} {\bibfnamefont
  {J.~M.}\ \bibnamefont {Gray}}, \bibinfo {author} {\bibfnamefont {H.~J.}\
  \bibnamefont {Hall}}, \bibinfo {author} {\bibfnamefont {V.~M.}\ \bibnamefont
  {Bright}}, \bibinfo {author} {\bibfnamefont {C.~T.}\ \bibnamefont {Rogers}},
  \bibinfo {author} {\bibfnamefont {M.~L.}\ \bibnamefont {Dunn}}, \bibinfo
  {author} {\bibfnamefont {R.~S.}\ \bibnamefont {Ruoff}},\ and\ \bibinfo
  {author} {\bibfnamefont {J.~S.}\ \bibnamefont {Bunch}},\ }\bibfield  {title}
  {\bibinfo {title} {Large arrays and properties of 3-terminal graphene
  nanoelectromechanical switches},\ }\href
  {https://doi.org/10.1002/adma.201304949} {\bibfield  {journal} {\bibinfo
  {journal} {Advanced Materials}\ }\textbf {\bibinfo {volume} {26}},\ \bibinfo
  {pages} {1571} (\bibinfo {year} {2013})}\BibitemShut {NoStop}%
\bibitem [{\citenamefont {Pereira}\ \emph {et~al.}(2009)\citenamefont
  {Pereira}, \citenamefont {Neto},\ and\ \citenamefont {Peres}}]{Pereira2009}%
  \BibitemOpen
  \bibfield  {author} {\bibinfo {author} {\bibfnamefont {V.~M.}\ \bibnamefont
  {Pereira}}, \bibinfo {author} {\bibfnamefont {A.~H.~C.}\ \bibnamefont
  {Neto}},\ and\ \bibinfo {author} {\bibfnamefont {N.~M.~R.}\ \bibnamefont
  {Peres}},\ }\bibfield  {title} {\bibinfo {title} {Tight-binding approach to
  uniaxial strain in graphene},\ }\href
  {https://doi.org/10.1103/physrevb.80.045401} {\bibfield  {journal} {\bibinfo
  {journal} {Phys. Rev. B}\ }\textbf {\bibinfo {volume} {80}},\ \bibinfo
  {pages} {045401} (\bibinfo {year} {2009})}\BibitemShut {NoStop}%
\bibitem [{\citenamefont {Ribeiro}\ \emph {et~al.}(2009)\citenamefont
  {Ribeiro}, \citenamefont {Pereira}, \citenamefont {Peres}, \citenamefont
  {Briddon},\ and\ \citenamefont {Neto}}]{Ribeiro2009}%
  \BibitemOpen
  \bibfield  {author} {\bibinfo {author} {\bibfnamefont {R.~M.}\ \bibnamefont
  {Ribeiro}}, \bibinfo {author} {\bibfnamefont {V.~M.}\ \bibnamefont
  {Pereira}}, \bibinfo {author} {\bibfnamefont {N.~M.~R.}\ \bibnamefont
  {Peres}}, \bibinfo {author} {\bibfnamefont {P.~R.}\ \bibnamefont {Briddon}},\
  and\ \bibinfo {author} {\bibfnamefont {A.~H.~C.}\ \bibnamefont {Neto}},\
  }\bibfield  {title} {\bibinfo {title} {Strained graphene: tight-binding and
  density functional calculations},\ }\href
  {https://doi.org/10.1088/1367-2630/11/11/115002} {\bibfield  {journal}
  {\bibinfo  {journal} {New Journal of Physics}\ }\textbf {\bibinfo {volume}
  {11}},\ \bibinfo {pages} {115002} (\bibinfo {year} {2009})}\BibitemShut
  {NoStop}%
\bibitem [{\citenamefont {Datta}(1997)}]{Datta1997}%
  \BibitemOpen
  \bibfield  {author} {\bibinfo {author} {\bibfnamefont {S.}~\bibnamefont
  {Datta}},\ }\href@noop {} {\emph {\bibinfo {title} {Electronic Transport in
  Mesoscopic Systems}}}\ (\bibinfo  {publisher} {Cambridge University Press},\
  \bibinfo {year} {1997})\BibitemShut {NoStop}%
\bibitem [{\citenamefont {Datta}(2005)}]{Datta2005}%
  \BibitemOpen
  \bibfield  {author} {\bibinfo {author} {\bibfnamefont {S.}~\bibnamefont
  {Datta}},\ }\href@noop {} {\emph {\bibinfo {title} {Quantum Transport: Atom
  to Transistor}}}\ (\bibinfo  {publisher} {Cambridge University Press},\
  \bibinfo {year} {2005})\BibitemShut {NoStop}%
\bibitem [{\citenamefont {Betancur-Ocampo}\ \emph {et~al.}(2019)\citenamefont
  {Betancur-Ocampo}, \citenamefont {Leyvraz},\ and\ \citenamefont
  {Stegmann}}]{Betancur2019}%
  \BibitemOpen
  \bibfield  {author} {\bibinfo {author} {\bibfnamefont {Y.}~\bibnamefont
  {Betancur-Ocampo}}, \bibinfo {author} {\bibfnamefont {F.}~\bibnamefont
  {Leyvraz}},\ and\ \bibinfo {author} {\bibfnamefont {T.}~\bibnamefont
  {Stegmann}},\ }\bibfield  {title} {\bibinfo {title} {Electron optics in
  phosphorene pn junctions: Negative reflection and anti-super-klein
  tunneling},\ }\href {https://doi.org/10.1021/acs.nanolett.9b02720} {\bibfield
   {journal} {\bibinfo  {journal} {Nano Lett.}\ }\textbf {\bibinfo {volume}
  {19}},\ \bibinfo {pages} {7760} (\bibinfo {year} {2019})}\BibitemShut
  {NoStop}%
\bibitem [{\citenamefont {Betancur-Ocampo}\ \emph {et~al.}(2020)\citenamefont
  {Betancur-Ocampo}, \citenamefont {Paredes-Rocha},\ and\ \citenamefont
  {Stegmann}}]{Betancur2020}%
  \BibitemOpen
  \bibfield  {author} {\bibinfo {author} {\bibfnamefont {Y.}~\bibnamefont
  {Betancur-Ocampo}}, \bibinfo {author} {\bibfnamefont {E.}~\bibnamefont
  {Paredes-Rocha}},\ and\ \bibinfo {author} {\bibfnamefont {T.}~\bibnamefont
  {Stegmann}},\ }\bibfield  {title} {\bibinfo {title} {Phosphorene pnp
  junctions as perfect electron waveguides},\ }\href
  {https://doi.org/10.1063/5.0019215} {\bibfield  {journal} {\bibinfo
  {journal} {Journal of Applied Physics}\ }\textbf {\bibinfo {volume} {128}},\
  \bibinfo {pages} {114303} (\bibinfo {year} {2020})}\BibitemShut {NoStop}%
\bibitem [{\citenamefont {Paredes-Rocha}\ \emph {et~al.}(2021)\citenamefont
  {Paredes-Rocha}, \citenamefont {Betancur-Ocampo}, \citenamefont {Szpak},\
  and\ \citenamefont {Stegmann}}]{Paredes2021}%
  \BibitemOpen
  \bibfield  {author} {\bibinfo {author} {\bibfnamefont {E.}~\bibnamefont
  {Paredes-Rocha}}, \bibinfo {author} {\bibfnamefont {Y.}~\bibnamefont
  {Betancur-Ocampo}}, \bibinfo {author} {\bibfnamefont {N.}~\bibnamefont
  {Szpak}},\ and\ \bibinfo {author} {\bibfnamefont {T.}~\bibnamefont
  {Stegmann}},\ }\bibfield  {title} {\bibinfo {title} {Gradient-index electron
  optics in graphene p-n junctions},\ }\bibfield  {journal} {\bibinfo
  {journal} {Physical Review B}\ }\textbf {\bibinfo {volume} {103}},\ \href
  {https://doi.org/10.1103/physrevb.103.045404} {10.1103/physrevb.103.045404}
  (\bibinfo {year} {2021})\BibitemShut {NoStop}%
\bibitem [{\citenamefont {de~Juan}\ \emph {et~al.}(2012)\citenamefont
  {de~Juan}, \citenamefont {Sturla},\ and\ \citenamefont
  {Vozmediano}}]{Juan2012}%
  \BibitemOpen
  \bibfield  {author} {\bibinfo {author} {\bibfnamefont {F.}~\bibnamefont
  {de~Juan}}, \bibinfo {author} {\bibfnamefont {M.}~\bibnamefont {Sturla}},\
  and\ \bibinfo {author} {\bibfnamefont {M.~A.~H.}\ \bibnamefont
  {Vozmediano}},\ }\bibfield  {title} {\bibinfo {title} {Space dependent fermi
  velocity in strained graphene},\ }\href
  {https://doi.org/10.1103/physrevlett.108.227205} {\bibfield  {journal}
  {\bibinfo  {journal} {Phys. Rev. Lett.}\ }\textbf {\bibinfo {volume} {108}},\
  \bibinfo {pages} {227205} (\bibinfo {year} {2012})}\BibitemShut {NoStop}%
\bibitem [{\citenamefont {de~Juan}\ \emph {et~al.}(2013)\citenamefont
  {de~Juan}, \citenamefont {Ma{\~{n}}es},\ and\ \citenamefont
  {Vozmediano}}]{Juan2013}%
  \BibitemOpen
  \bibfield  {author} {\bibinfo {author} {\bibfnamefont {F.}~\bibnamefont
  {de~Juan}}, \bibinfo {author} {\bibfnamefont {J.~L.}\ \bibnamefont
  {Ma{\~{n}}es}},\ and\ \bibinfo {author} {\bibfnamefont {M.~A.~H.}\
  \bibnamefont {Vozmediano}},\ }\bibfield  {title} {\bibinfo {title} {Gauge
  fields from strain in graphene},\ }\href
  {https://doi.org/10.1103/physrevb.87.165131} {\bibfield  {journal} {\bibinfo
  {journal} {Phys. Rev. B}\ }\textbf {\bibinfo {volume} {87}},\ \bibinfo
  {pages} {165131} (\bibinfo {year} {2013})}\BibitemShut {NoStop}%
\bibitem [{\citenamefont {Oliva-Leyva}\ and\ \citenamefont
  {Naumis}(2015)}]{Oliva-Leyva2015}%
  \BibitemOpen
  \bibfield  {author} {\bibinfo {author} {\bibfnamefont {M.}~\bibnamefont
  {Oliva-Leyva}}\ and\ \bibinfo {author} {\bibfnamefont {G.~G.}\ \bibnamefont
  {Naumis}},\ }\bibfield  {title} {\bibinfo {title} {Generalizing the fermi
  velocity of strained graphene from uniform to nonuniform strain},\ }\href
  {https://doi.org/10.1016/j.physleta.2015.05.039} {\bibfield  {journal}
  {\bibinfo  {journal} {Phys. Lett. A}\ }\textbf {\bibinfo {volume} {379}},\
  \bibinfo {pages} {2645} (\bibinfo {year} {2015})}\BibitemShut {NoStop}%
\bibitem [{\citenamefont {Castro~Neto}\ \emph {et~al.}(2009)\citenamefont
  {Castro~Neto}, \citenamefont {Guinea}, \citenamefont {Peres}, \citenamefont
  {Novoselov},\ and\ \citenamefont {Geim}}]{CastroNeto2009}%
  \BibitemOpen
  \bibfield  {author} {\bibinfo {author} {\bibfnamefont {A.~H.}\ \bibnamefont
  {Castro~Neto}}, \bibinfo {author} {\bibfnamefont {F.}~\bibnamefont {Guinea}},
  \bibinfo {author} {\bibfnamefont {N.~M.~R.}\ \bibnamefont {Peres}}, \bibinfo
  {author} {\bibfnamefont {K.~S.}\ \bibnamefont {Novoselov}},\ and\ \bibinfo
  {author} {\bibfnamefont {A.~K.}\ \bibnamefont {Geim}},\ }\bibfield  {title}
  {\bibinfo {title} {The electronic properties of graphene},\ }\href
  {https://doi.org/10.1103/RevModPhys.81.109} {\bibfield  {journal} {\bibinfo
  {journal} {Rev. Mod. Phys.}\ }\textbf {\bibinfo {volume} {81}},\ \bibinfo
  {pages} {109} (\bibinfo {year} {2009})}\BibitemShut {NoStop}%
\bibitem [{\citenamefont {Cooper}\ \emph {et~al.}(2012)\citenamefont {Cooper},
  \citenamefont {D’Anjou}, \citenamefont {Ghattamaneni}, \citenamefont
  {Harack}, \citenamefont {Hilke}, \citenamefont {Horth}, \citenamefont
  {Majlis}, \citenamefont {Massicotte}, \citenamefont {Vandsburger},
  \citenamefont {Whiteway} \emph {et~al.}}]{Cooper2012}%
  \BibitemOpen
  \bibfield  {author} {\bibinfo {author} {\bibfnamefont {D.~R.}\ \bibnamefont
  {Cooper}}, \bibinfo {author} {\bibfnamefont {B.}~\bibnamefont {D’Anjou}},
  \bibinfo {author} {\bibfnamefont {N.}~\bibnamefont {Ghattamaneni}}, \bibinfo
  {author} {\bibfnamefont {B.}~\bibnamefont {Harack}}, \bibinfo {author}
  {\bibfnamefont {M.}~\bibnamefont {Hilke}}, \bibinfo {author} {\bibfnamefont
  {A.}~\bibnamefont {Horth}}, \bibinfo {author} {\bibfnamefont
  {N.}~\bibnamefont {Majlis}}, \bibinfo {author} {\bibfnamefont
  {M.}~\bibnamefont {Massicotte}}, \bibinfo {author} {\bibfnamefont
  {L.}~\bibnamefont {Vandsburger}}, \bibinfo {author} {\bibfnamefont
  {E.}~\bibnamefont {Whiteway}}, \emph {et~al.},\ }\bibfield  {title} {\bibinfo
  {title} {Experimental review of graphene},\ }\bibfield  {journal} {\bibinfo
  {journal} {International Scholarly Research Notices}\ }\textbf {\bibinfo
  {volume} {2012}},\ \href {https://doi.org/10.5402/2012/501686}
  {10.5402/2012/501686} (\bibinfo {year} {2012})\BibitemShut {NoStop}%
\bibitem [{\citenamefont {Shi}\ \emph {et~al.}(2019)\citenamefont {Shi},
  \citenamefont {Fan}, \citenamefont {Li},\ and\ \citenamefont {Li}}]{Shi2019}%
  \BibitemOpen
  \bibfield  {author} {\bibinfo {author} {\bibfnamefont {F.-T.}\ \bibnamefont
  {Shi}}, \bibinfo {author} {\bibfnamefont {S.-C.}\ \bibnamefont {Fan}},
  \bibinfo {author} {\bibfnamefont {C.}~\bibnamefont {Li}},\ and\ \bibinfo
  {author} {\bibfnamefont {Z.-A.}\ \bibnamefont {Li}},\ }\bibfield  {title}
  {\bibinfo {title} {Opto-thermally excited fabry-perot resonance frequency
  behaviors of clamped circular graphene membrane},\ }\href
  {https://doi.org/10.3390/nano9040563} {\bibfield  {journal} {\bibinfo
  {journal} {Nanomaterials}\ }\textbf {\bibinfo {volume} {9}},\ \bibinfo
  {pages} {563} (\bibinfo {year} {2019})}\BibitemShut {NoStop}%
\end{thebibliography}%

\end{document}